\documentclass[11pt,a4paper]{article}
\usepackage[T1]{fontenc}
\usepackage[utf8]{inputenc}
\usepackage{lmodern}
\usepackage[a4paper,left=26mm,right=26mm,top=28mm,bottom=27mm,includefoot,footskip=18pt]{geometry}
\usepackage{amsmath,amssymb,amsthm,mathtools}
\usepackage{enumitem}
\usepackage{tikz}
\usetikzlibrary{decorations.pathreplacing}
\usepackage[title]{appendix}
\usepackage{placeins}
\usepackage[colorlinks=true,linkcolor=black,citecolor=black,urlcolor=black,bookmarksnumbered=true]{hyperref}
\hypersetup{pdftitle={The Functional Batch Conjecture And Its Extension To All Finite Fields},
  pdfauthor={Isaac Barouch Essayag, Aryeh Lev Zabokritskiy (Yohananov)},pdflang={en}}
\newtheorem{theorem}{Theorem}[section]
\newtheorem{proposition}[theorem]{Proposition}
\newtheorem{lemma}[theorem]{Lemma}
\newtheorem{corollary}[theorem]{Corollary}
\theoremstyle{definition}
\newtheorem{definition}[theorem]{Definition}
\newtheorem{example}[theorem]{Example}
\theoremstyle{remark}
\newtheorem{remark}[theorem]{Remark}

\newcommand{\F}{\mathbb{F}}
\newcommand{\Z}{\mathbb{Z}}
\newcommand{\Q}{\mathbb{Q}}
\newcommand{\C}{\mathbb{C}}
\newcommand{\R}{\mathbb{R}}
\newcommand{\cP}{\mathcal{P}}
\newcommand{\cO}{\mathcal{O}}
\newcommand{\cW}{\mathcal{W}}
\newcommand{\cT}{\mathcal{T}}
\newcommand{\cA}{\mathcal{A}}
\newcommand{\fm}{\mathfrak{m}}
\newcommand{\fB}{\mathfrak{B}}
\newcommand{\Pf}{\operatorname{Pf}}
\newcommand{\tr}{\operatorname{tr}}
\newcommand{\sgn}{\operatorname{sgn}}
\newcommand{\hgt}{\operatorname{lev}}
\newcommand{\Sw}{\operatorname{Sw}}
\newcommand{\FB}{\operatorname{FB}}
\newcommand{\FP}{\operatorname{FP}}
\newcommand{\spn}[1]{\langle #1\rangle}
\newcommand{\Tqs}{T_{q,s}}
\newcommand{\Gqs}{G_{q,s}}

\title{The Functional Batch Conjecture\\
And Its Extension To All Finite Fields}
\author{%
  \normalsize Isaac Barouch Essayag\\[0.45em]
  \small MIGAL--Galilee Research Institute, Kiryat Shmona, Israel\\
  \small Tel-Hai University of Kiryat Shmona in the Galilee, Israel\\[0.25em]
  \small\href{mailto:isaac.es@migal.org.il}{\texttt{isaac.es@migal.org.il}}\\[1.15em]
  \normalsize Aryeh Lev Zabokritskiy (Yohananov)\\[0.45em]
  \small Department of Computer Science, MIGAL--Galilee Research Institute,\\
  \small Kiryat Shmona, Israel, and Department of Computer Science, Tel-Hai\\
  \small University of Kiryat Shmona in the Galilee, Kiryat Shmona, Israel\\[0.25em]
  \small\href{mailto:yuhanalev@telhai.ac.il}{\texttt{yuhanalev@telhai.ac.il}}\\[0.35em]
  \small ORCID: \href{https://orcid.org/0000-0003-3151-6192}{0000-0003-3151-6192}%
}
\date{}
\begin{document}
\maketitle

\begin{abstract}
We prove that the code whose servers store all $q^s-1$ nonzero linear combinations of $s$
information symbols over $\F_q$ is a functional batch code with list size $(q^s+q-2)/2$: every
list of $(q^s+q-2)/2$ requested nonzero linear combinations, repetitions allowed, can be recovered
from pairwise disjoint sets of at most two servers. This holds for every finite field and every
$s\ge1$, and the list size cannot be increased, even if recovery sets of any size are allowed.
For $q=2$ the statement is the functional batch
conjecture for the binary simplex code, in its stronger form with recovery sets of size at most two, and it
follows from the conjecture of Balister, Gy\H{o}ri and Schelp on perfect matchings of $\F_2^n$
with prescribed differences, which was proved by the second author in a separate paper. We also
determine the minimum length of a functional batch code with $s$ information symbols over
$\F_q$: it is $m(q^s-1)$ for the list size $m(q^s+q-2)/2$, for every $m\ge1$, and, for odd $q$,
it is $2(q^s-1)/(q-1)$ for the list size $(q^s-1)/(q-1)+1$.
\end{abstract}

\section{Introduction}\label{sec:intro}

Let $q$ be a prime power and $s\ge1$. Given a list $r_1,\dots,r_t$ of nonzero vectors of
$\F_q^s$, in which repetitions are allowed, we look for pairwise disjoint sets $R_1,\dots,R_t$ of
nonzero vectors of $\F_q^s$, each of size one or two, such that $r_i$ lies in the span of $R_i$
for every $i$. For which $t$ can this be done for every list? We show that it can be done exactly
when $t\le(q^s+q-2)/2$; for $q=2$ the case $t=2^{s-1}$ was conjectured in~\cite{HKRS} and follows
from~\cite{Bin}, as recalled below. In the language of batch codes the $q^s-1$ nonzero vectors
are servers, the $r_i$ are requests and the $R_i$ are recovery sets, and the
result says that these servers form a functional batch code with this list size in which every
request is recovered from at most two servers.

Batch codes were introduced by Ishai, Kushilevitz, Ostrovsky and Sahai~\cite{IKOS} for load
balancing in distributed storage, and PIR codes by Fazeli, Vardy and Yaakobi~\cite{FVY} in order
to reduce the storage overhead of private information retrieval. In linear batch and PIR
codes, $s$ information symbols from a finite field $\F_q$ are stored on $n$ servers, and the
$j$th server holds the linear combination of the symbols whose coefficient vector is the $j$th
column of an $s\times n$ matrix $G$ over $\F_q$. In the functional versions of these codes, a
\emph{request} is any nonzero vector
$r\in\F_q^s$; it stands for the linear combination of the information symbols that a user asks
for. A set $R$ of servers is a \emph{recovery set} for $r$ if $r$ lies in the span of the
corresponding columns, and $G$ \emph{serves} a list $r_1,\dots,r_t$ of requests if these requests
have pairwise disjoint recovery sets, so that $t$ users are answered at the same time, each from
its own servers. Zhang, Etzion and Yaakobi~\cite{ZEY} studied matrices that serve every list of
$t$ requests; following~\cite{KRS} we call such a matrix a \emph{$t$-functional batch code}, and we call a matrix a
\emph{$t$-functional PIR code} if it serves every list that consists of $t$ copies of one request.
A list may contain repeated requests.

The best studied example is the binary simplex code of dimension $s$, whose generator matrix has
the $2^s-1$ nonzero vectors of $\F_2^s$ as its columns. Wang, Kiah, Cassuto and Bruck~\cite{WKCB}
proved that this matrix serves every list of $2^{s-1}$ unit vectors, and Zhang, Etzion and
Yaakobi conjectured that it is a $2^{s-1}$-functional batch code~\cite[Conjecture~25]{ZEY}, a
statement called
the functional batch conjecture in~\cite{KRS}.
Hollmann, Khathuria, Riet and Skachek~\cite[Section~2]{HKRS} conjectured more: every list of
$2^{s-1}$ requests can be served with recovery sets of size at most two. Following~\cite{OHRS},
we say that a matrix serves a list \emph{with locality two} if it serves it with recovery sets of
size at most two. A set $\{x,y\}$ of two columns of the generator
matrix of the binary simplex code recovers the request $x+y$. For this reason, and by a short reduction for lists
whose sum is not zero, both conjectures
follow from a conjecture of Balister, Gy\H{o}ri and Schelp~\cite{BGS} on perfect matchings: if
$n\ge2$, then every list of $2^{n-1}$ nonzero vectors of $\F_2^n$ with sum zero is the list of
differences of a perfect matching of $\F_2^n$. This connection is described by Yohananov and
Essayag~\cite{YE} for the conjecture of~\cite{ZEY}, and it is pointed out in~\cite{KRS} for the
conjecture of~\cite{HKRS}. Several families of lists are known to be served by
the binary simplex code. Yohananov and Yaakobi~\cite{YY} showed that it serves every list of
$\lfloor\frac56 2^{s-1}\rfloor-s$ requests; for every hyperplane, the lists of $2^{s-1}$ requests
that lie outside it are served with locality two~\cite{HKRS}, by an argument that goes back to a
theorem of M.~Hall~\cite{Hall} on abelian groups. Further special
families of lists are treated in~\cite{BGNY,Erde,Kov,KZ,Zab}, and it is noted in~\cite{KRS} that a
rainbow matching theorem of~\cite{CPS} implies that the binary simplex code serves every list
of $2^{s-1}-O(2^{15s/16})$ requests when $s$ is large. Finally, the conjecture of Balister,
Gy\H{o}ri and Schelp is proved in~\cite{Bin}; consequently the two conjectures on the binary
simplex code hold.

Much less is known over other fields. K\i l\i\c{c}, Ravagnani and Salizzoni~\cite{KRS} studied
the minimum length $\FB(s,t,q)$ of a $t$-functional batch code with $s$ information symbols over
$\F_q$, and the corresponding quantity $\FP(s,t,q)$ for functional PIR codes. For functional PIR
codes they identify the counterparts of the generator matrix of the binary simplex code and of
the list size $2^{s-1}$:
the $s\times(q^s-1)$ matrix $\Gqs$ whose columns are all nonzero vectors of $\F_q^s$, and the
list size
\[
  \Tqs \coloneqq \frac{q^s+q-2}{2}.
\]
They prove that $\Gqs$ serves $\Tqs$ copies of every request and that no matrix with fewer
columns has this property, so that $\FP(s,\Tqs,q)=q^s-1$ \cite[Lemma~3.10]{KRS}, and more
generally that $\FP(s,m\Tqs,q)=m(q^s-1)$ for every $m\ge1$ \cite[Theorem~3.11]{KRS}. For lists of
arbitrary requests they pose two open problems on the function
$\FB$~\cite[Open Problems~3.12 and~3.13]{KRS}.

Our main result is the following.

\begin{theorem}\label{thm:main}
Let $q$ be a prime power and $s\ge1$, and recall that $\Tqs=(q^s+q-2)/2$.
\begin{enumerate}[label=\textup{(\alph*)}]
\item\label{main:a} For every list $r_1,\dots,r_{\Tqs}$ of nonzero vectors of $\F_q^s$ there are
pairwise disjoint sets $R_1,\dots,R_{\Tqs}$ of nonzero vectors of $\F_q^s$, each of size one or
two, such that every $r_i$ is a linear combination with nonzero coefficients of the elements of
$R_i$.
\item\label{main:b} If $v\in\F_q^s$ is nonzero, then no list of $\Tqs+1$ nonzero multiples of $v$
is served by $\Gqs$, even when recovery sets of arbitrary size are allowed.
\end{enumerate}
\end{theorem}

In part~\ref{main:a} the requirement that the coefficients are nonzero is no restriction, because
a server whose coefficient is zero can be removed from $R_i$. Thus $\Gqs$ is a $\Tqs$-functional
batch code with locality two, and by part~\ref{main:b} $\Tqs$ is the largest list size for which
this holds. The matrix $\Gqs$ and the list size $\Tqs$ are the ones for which it is shown
in~\cite{KRS} that $\Tqs$ copies of every request are served; by Theorem~\ref{thm:main} the same
matrix serves every list of $\Tqs$ requests, with pairwise disjoint recovery sets of size at most
two.
Part~\ref{main:a} applies to every list: requests may be repeated, and no condition is imposed on
their sum. For $q=2$ part~\ref{main:a} is the conjecture of~\cite{HKRS}, and it
implies the conjecture of~\cite{ZEY}. Part~\ref{main:b} follows from the counting argument that is used
in~\cite{KRS} for copies of one request; the same argument shows that every matrix that serves
$t$ copies of every request has at least $2t(q^s-1)/(q^s+q-2)$ columns.

Let $N\coloneqq(q^s-1)/(q-1)$ denote the number of one-dimensional subspaces of $\F_q^s$. For odd
$q$ we prove a stronger statement: if any two distinct nonzero vectors
are chosen in each
one-dimensional subspace of $\F_q^s$, then the
resulting $2N$ servers serve every list of $N+1$ requests with locality two
(Theorem~\ref{thm:tworep}). It implies
Theorem~\ref{thm:main}\ref{main:a} for these $q$, because $\Tqs=\frac{q-1}{2}(N+1)$ and the $q-1$
nonzero vectors of each one-dimensional subspace split into $(q-1)/2$ pairs.
Theorem~\ref{thm:main} concerns the single matrix $\Gqs$; together with this statement on $2N$
servers it has the following consequences for the minimum length over all matrices.

\begin{corollary}\label{cor:length}
Let $q$ be a prime power and $s\ge1$.
\begin{enumerate}[label=\textup{(\alph*)}]
\item\label{length:a} $\FB(s,m\Tqs,q)=m(q^s-1)$ for every integer $m\ge1$.
\item\label{length:b} If $q$ is odd, then $\FB(s,N+1,q)=2N$.
\end{enumerate}
\end{corollary}

Open Problems~3.12 and~3.13 of~\cite{KRS} ask whether $\FB(s,2\Tqs,q)=2(q^s-1)$ and whether
$\FB(s,N+1,q)=2N$; for $q=2$ both ask for the value $\FB(s,2^s,2)=2(2^s-1)$, which is
known~\cite{HKRS,KRS,YY}, and for every $q$ it is shown in~\cite[Proposition~4.11]{KRS} that
$\FB(s,2\Tqs,q)\le2(q^s-1)+(q-2)s$. Corollary~\ref{cor:length}\ref{length:a} with $m=2$ answers
the first in the affirmative, and Corollary~\ref{cor:length}\ref{length:b}, which rests on the
statement on $2N$ servers, answers the second in the affirmative for odd~$q$; we do not know whether the second holds for even $q\ge4$ and $s\ge3$.

Some cases of Theorem~\ref{thm:main}\ref{main:a} and of Corollary~\ref{cor:length} are known.
For $s=1$ they are immediate. For $s=2$, \cite[Theorem~3.1]{KRS} determines $\FB(2,t,q)$ for all
$t$ and $q$; this contains both parts of Corollary~\ref{cor:length}, and we show that
Theorem~\ref{thm:main}\ref{main:a} and, for odd $q$, the statement on $2N$ servers can be deduced
from it.
For $q=2$ and $s\le5$, Theorem~\ref{thm:main}\ref{main:a} follows from the verification of the
conjecture of~\cite{BGS} for $n\le5$ by Balister, Gy\H{o}ri and Schelp, which is reported
in~\cite{Kov}; for
$3\le s\le5$ the values $\FB(s,2^{s-1},2)=2^s-1$ are recorded in~\cite{ZEY}, and together with
the counting argument of~\cite{KRS} they give Corollary~\ref{cor:length}\ref{length:a} for
every~$m$. For $q=2$ and even $m$,
Corollary~\ref{cor:length}\ref{length:a} is \cite[Lemma~3.6]{KRS}. For $q=2$ and every $s\ge2$,
Theorem~\ref{thm:main}\ref{main:a} follows by a short reduction from the conjecture
of~\cite{BGS}, which is proved in~\cite{Bin}, and with the counting argument this gives
Corollary~\ref{cor:length}\ref{length:a} for $q=2$. In all other cases, that is, for $q\ge3$ and
$s\ge3$, the lower bounds in Corollary~\ref{cor:length} come from this counting argument, and our
contribution is the existence of the recovery sets.

The conjecture of Balister, Gy\H{o}ri and Schelp concerns perfect matchings of the
whole group $\F_2^n$: the list has exactly $2^{n-1}$ entries and sum zero, the zero vector is a
vertex, and every edge is a pair. In Theorem~\ref{thm:main}\ref{main:a} the sum of the list is
arbitrary, the zero vector is not a server, and a recovery set may be a single server; for $q=2$
the short reduction mentioned above passes from this conjecture to
Theorem~\ref{thm:main}\ref{main:a}. For $q>2$ a recovery set $\{x,y\}$ recovers every vector
$ax+by$ with nonzero coefficients $a$ and $b$, whereas a prescribed difference fixes the
coefficients; Theorem~\ref{thm:main}\ref{main:a} is therefore a statement about disjoint recovery
sets and not about matchings with prescribed differences.

For even $q$ our proof of Theorem~\ref{thm:main}\ref{main:a} uses as an input the conjecture of
Balister, Gy\H{o}ri and Schelp (Theorem~\ref{thm:bgs}), which is proved in~\cite{Bin}; so that
the present
paper can be read independently, we include a complete proof of it, which follows~\cite{Bin}. All our
proofs of Theorem~\ref{thm:main}\ref{main:a} establish the existence of the recovery sets, and an
efficient procedure that finds them is left open.

The paper is organized as follows. Section~\ref{sec:model} fixes the model, proves
Theorem~\ref{thm:main}\ref{main:b}, derives Corollary~\ref{cor:length} from
Theorem~\ref{thm:main}\ref{main:a} and the statement on $2N$ servers, and outlines the proof of
Theorem~\ref{thm:main}\ref{main:a}, which is different for odd and for even $q$.
Section~\ref{sec:odd} treats odd $q$ and Section~\ref{sec:even} treats even $q$.
Section~\ref{sec:conclusion} contains concluding remarks.

\section{Model, statements and outline of the proofs}\label{sec:model}

\subsection{Servers, requests and recovery sets}\label{subsec:model}

Throughout, $q$ is a prime power, $s\ge1$, and $V\coloneqq\F_q^s$. A \emph{line} is a
one-dimensional subspace of $V$; we write $\spn{v}$ for the line spanned by a nonzero vector $v$,
and $\cP$ for the set of lines; their number is $N=(q^s-1)/(q-1)$. The nonzero vectors of $V$ are
the columns of $\Gqs$, and we identify each of them with the server that holds it.

\begin{definition}\label{def:serve}
Let $S$ be a set of nonzero vectors of $V$, called \emph{servers}. A \emph{request} is a nonzero
vector $r\in V$. A subset $R\subseteq S$ is a \emph{recovery set} for $r$ if $r$ is a linear
combination of the elements of $R$. The set $S$ \emph{serves} a list $r_1,\dots,r_t$ of requests
if there are pairwise disjoint subsets $R_1,\dots,R_t$ of $S$ such that $R_i$ is a recovery set
for $r_i$ for every $i$. It serves the list \emph{with locality two} if in addition every $R_i$ has
one or two elements and $r_i$ is a linear combination of the elements of $R_i$ in which all
coefficients are nonzero.
\end{definition}

For $S=V\setminus\{0\}$, serving by $S$ is serving by the matrix $\Gqs$ in the sense of the
Introduction, and Theorem~\ref{thm:main}\ref{main:a} states that $V\setminus\{0\}$ serves every
list of $\Tqs$ requests with locality two. If $R$ is a recovery set for $r$ with at most two
elements, then $r$ is a linear combination with nonzero coefficients of the elements of some
nonempty subset of $R$. Hence $S$ serves a list with locality two if and only if it serves the
list with recovery sets of size at most two, as in the notion of locality two
of~\cite{OHRS}, which imposes this on every list of a given size. A recovery set of one element for $r$ consists of a server on the line
$\spn{r}$. The order of a list is irrelevant, and a request may occur several times. A request
$r$ and a multiple $\lambda r$ with $\lambda\in\F_q\setminus\{0\}$ have the same recovery sets, and
a linear combination with nonzero coefficients for one of them gives one for the other. We
therefore say that a request $r$ \emph{lies on} the line $\spn{r}$. In a list we may replace any
request by a nonzero multiple of it, so whether $S$ serves a list, with or without locality two,
depends only on the number of requests on each line.

\subsection{The upper bound}\label{subsec:upper}

Lists whose requests all lie on one line bound the number $t$ for which a set of servers serves
every list of $t$ requests. The proof of the next proposition is the counting argument used in
the proof of~\cite[Lemma~3.10]{KRS} for copies of one request.

\begin{proposition}\label{prop:upper}
Let $S$ be a set of nonzero vectors of $V$ and let $v\in V$ be nonzero. If $S$ serves a list of
$t$ requests on the line $\spn{v}$, with recovery sets of arbitrary size, then
\[
  t\;\le\;|S\cap\spn{v}|+\Bigl\lfloor\tfrac12\,|S\setminus\spn{v}|\Bigr\rfloor .
\]
\end{proposition}

\begin{proof}
Let $R_1,\dots,R_t$ be pairwise disjoint recovery sets for the $t$ requests. At most
$|S\cap\spn{v}|$ of them contain an element of $\spn{v}$. Each of the others is contained in
$S\setminus\spn{v}$ and has at least two elements, because a single vector $u\notin\spn{v}$ spans
the line $\spn{u}\ne\spn{v}$. Hence there are at most $\lfloor|S\setminus\spn{v}|/2\rfloor$ of the
latter.
\end{proof}

\begin{proof}[Proof of Theorem~\ref{thm:main}\ref{main:b}]
For $S=V\setminus\{0\}$ we have $|S\cap\spn{v}|=q-1$ and $|S\setminus\spn{v}|=q^s-q$, which is
even. Proposition~\ref{prop:upper} gives $t\le (q-1)+(q^s-q)/2=\Tqs$.
\end{proof}

\subsection{Two servers on each line for odd \texorpdfstring{$q$}{q}}\label{subsec:tworep}

For odd $q$ we prove Theorem~\ref{thm:main}\ref{main:a} through the following statement about
$2N$ servers.

\begin{theorem}\label{thm:tworep}
Let $q$ be odd. For every line $L\in\cP$ choose two distinct nonzero vectors of $L$, and let $S$
be the set of the $2N$ chosen vectors. Then $S$ serves every list of $N+1$ requests with locality
two.
\end{theorem}

By Proposition~\ref{prop:upper}, a set $S$ as in Theorem~\ref{thm:tworep} serves at most
$2+(2N-2)/2=N+1$ requests on one line, so the list size $N+1$ in Theorem~\ref{thm:tworep} is the
largest possible. Theorem~\ref{thm:tworep} is proved in Section~\ref{sec:odd}. It implies the main
theorem for odd $q$ at once.

\begin{proof}[Proof of Theorem~\ref{thm:main}\ref{main:a} for odd $q$, assuming Theorem~\ref{thm:tworep}]
Each line contains $q-1$ nonzero vectors, and $q-1$ is even. Partition the nonzero vectors of
every line into $(q-1)/2$ pairs, and number the pairs of each line from $1$ to $(q-1)/2$. For
$1\le j\le(q-1)/2$ let $S_j$ be the union over all lines of the $j$th pairs. The sets $S_j$ are
pairwise disjoint, and each of them consists of two nonzero vectors on every line. Since
\[
  \frac{q-1}{2}\,(N+1)=\frac{(q^s-1)+(q-1)}{2}=\Tqs,
\]
a list of $\Tqs$ requests splits into $(q-1)/2$ lists of $N+1$ requests. By
Theorem~\ref{thm:tworep} the $j$th of these lists is served with locality two by $S_j$.
\end{proof}

\subsection{The matching theorem used for even \texorpdfstring{$q$}{q}}\label{subsec:input}

For even $q$ our proof of Theorem~\ref{thm:main}\ref{main:a} rests on the following statement
about the group $\F_2^n$.

\begin{theorem}\label{thm:bgs}
Let $n\ge2$ and let $d_1,\dots,d_{2^{n-1}}$ be nonzero vectors of $\F_2^n$ whose sum is zero. Then
there are vectors $x_1,\dots,x_{2^{n-1}}$ such that the pairs $\{x_i,x_i+d_i\}$,
$1\le i\le2^{n-1}$, form a partition of $\F_2^n$.
\end{theorem}

Theorem~\ref{thm:bgs} was conjectured by Balister, Gy\H{o}ri and Schelp~\cite{BGS}, who verified
it for $n\le5$, as reported in~\cite{Kov}. It is proved in~\cite[Theorem~1.1]{Bin}. So that the
present paper
can be read independently, Appendix~\ref{app:bgs} contains a complete proof, which
follows~\cite{Bin}. The proof counts the partitions in question with signs and
with the labellings of their pairs, shows that this count is a coefficient of a Pfaffian, and
shows that the count is divisible by $2^{2^{n-1}-1}$ and not by $2^{2^{n-1}}$, so that it is not
zero. This proof uses no other result of this paper, and Sections~\ref{sec:odd}
and~\ref{sec:even} can be read without it. In the proof of Theorem~\ref{thm:main}\ref{main:a},
Theorem~\ref{thm:bgs} is the main ingredient for $q=2$, and for even $q\ge4$ it is applied to one
subgroup of $V$. For $q=2$ the proof uses Theorem~\ref{thm:bgs} only with $n=s$
(Subsection~\ref{subsec:inputs}), so the cases $s\le5$ of
Theorem~\ref{thm:main}\ref{main:a} for $q=2$ follow from the verification mentioned above.

\subsection{Consequences for the minimum length}\label{subsec:length}

We recall the definitions of~\cite{KRS}, where the number of information symbols is denoted by
$k$. Let $M$ be an $s\times n$ matrix over $\F_q$; its columns may be repeated and may be zero.
A set $R$ of
column indices is a recovery set for a request $r$ if $r$ lies in the span of the columns with
index in $R$, and $M$ serves a list of requests if the requests have
pairwise disjoint recovery sets~\cite[Definition~2.2]{KRS}. The number $\FB(s,t,q)$ is the
smallest $n$ for which some $s\times n$ matrix over $\F_q$ serves every list of $t$ requests, and
$\FP(s,t,q)$ is the smallest $n$ for which some $s\times n$ matrix serves, for every request $r$,
the list of $t$ copies of $r$ \cite[Definitions~2.3 and~2.4]{KRS}. Clearly
$\FP(s,t,q)\le\FB(s,t,q)$. The lower bounds that we need follow from the next lemma, whose proof
is the counting argument in the proofs of \cite[Lemma~3.10 and Theorem~3.11]{KRS}.

\begin{lemma}\label{lem:counting}
If an $s\times n$ matrix over $\F_q$ serves, for every request $r$, the list of $t$ copies of $r$,
then
\[
  n\;\ge\;\frac{2tN}{N+1}=\frac{2t\,(q^s-1)}{q^s+q-2}.
\]
\end{lemma}

\begin{proof}
For a line $L$ let $n_L$ be the number of column indices whose column is a nonzero vector of $L$.
Then $\sum_{L\in\cP}n_L\le n$, so some line $L$ satisfies $n_L\le n/N$. Let $v$ be a nonzero
vector of $L$, and consider pairwise disjoint recovery sets for $t$ copies of $v$. At most $n_L$
of them contain the index of a column in $L\setminus\{0\}$. Each of the others consists of indices
of columns outside $L\setminus\{0\}$ and has at least two elements, since neither the zero vector
nor a single vector outside $L$ spans $v$. Therefore
$t\le n_L+(n-n_L)/2\le\frac12\,n\,(1+1/N)$, which is the assertion. The second expression for the
bound follows from $N+1=(q^s+q-2)/(q-1)$.
\end{proof}

\begin{proof}[Proof of Corollary~\ref{cor:length}, assuming Theorems~\ref{thm:main}\ref{main:a} and~\ref{thm:tworep}]
\ref{length:a} Let $M$ be the $s\times m(q^s-1)$ matrix that consists of $m$ copies of $\Gqs$
side by side. A list of $m\Tqs$ requests splits into $m$ lists of $\Tqs$ requests, and by
Theorem~\ref{thm:main}\ref{main:a} the $j$th of these lists is served by the $j$th copy. Hence
$\FB(s,m\Tqs,q)\le m(q^s-1)$. Conversely, Lemma~\ref{lem:counting} with $t=m\Tqs$ and
$2\Tqs=q^s+q-2$ gives $\FB(s,m\Tqs,q)\ge\FP(s,m\Tqs,q)\ge m(q^s-1)$.

\ref{length:b} The matrix whose columns are the $2N$ vectors of a set $S$ as in
Theorem~\ref{thm:tworep} serves every list of $N+1$ requests, so $\FB(s,N+1,q)\le2N$.
Lemma~\ref{lem:counting} with $t=N+1$ gives $\FB(s,N+1,q)\ge\FP(s,N+1,q)\ge2N$.
\end{proof}

The equality $\FP(s,m\Tqs,q)=m(q^s-1)$, which the proof of part~\ref{length:a} contains, is
\cite[Theorem~3.11]{KRS}, stated there with $k$ and $s$ in place of our $s$ and $m$. Some cases
of the corollary itself are known. For $s\le2$ both parts are contained in~\cite{KRS}:
$\FB(1,t,q)=t$, and \cite[Theorem~3.1]{KRS} states that
$\FB(2,t,q)=\lceil 2(q+1)t/(q+2)\rceil$ for all $t$, which is $m(q^2-1)$ for $t=mT_{q,2}$ and
$2(q+1)$ for $t=q+2$. For $q=2$ and even $m$, part~\ref{length:a} is the second equality of
\cite[Lemma~3.6]{KRS}, stated there with $k$ and $s$ in place of our $s$ and $m/2$. For $q=2$ and
$3\le s\le5$, the case $m=1$ of part~\ref{length:a} is recorded in the discussion before
\cite[Conjecture~25]{ZEY}, and the other values of $m$ follow from it as in the proof above,
with $m$ copies of a matrix that attains this value. For $q=2$ and every $s\ge2$,
Theorem~\ref{thm:main}\ref{main:a} is a consequence of Theorem~\ref{thm:bgs}
(Lemma~\ref{lem:binary}), hence of~\cite{Bin}, and part~\ref{length:a} follows from it as in the
proof above.

For $s=2$, Theorem~\ref{thm:main}\ref{main:a} can be deduced from~\cite[Theorem~3.1]{KRS}. By
that theorem there is a matrix with $q^2-1$ columns that serves every list of $T_{q,2}$ requests.
For every line $L$, let $n_L$ be the number of columns of such a matrix that are nonzero vectors
of $L$; the count in the proof of Lemma~\ref{lem:counting} gives
$T_{q,2}\le n_L+(q^2-1-n_L)/2$, that is, $n_L\ge q-1$. Since there are $q+1$ lines, such a matrix
has exactly $q-1$ nonzero columns on each line. Multiplying the columns on each line by suitable
nonzero scalars turns them into the $q-1$ nonzero vectors of that line and does not change the
span of any set of columns, so $G_{q,2}$ also serves every list of $T_{q,2}$ requests. Finally,
in $\F_q^2$ every recovery set contains a recovery set with at most two elements, so these lists
are served with recovery sets of size at most two, that is, with locality two
(Subsection~\ref{subsec:model}). In the same way, the value $\FB(2,q+2,q)=2(q+1)$ gives a matrix
with exactly two nonzero columns on each line that serves every list of $q+2$ requests, and
Theorem~\ref{thm:tworep} for $s=2$ follows. This deduction is valid for every $q\ge3$.

\subsection{Outline of the proof of Theorem~\ref{thm:main}\ref{main:a}}\label{subsec:outline}

For $s=1$ the theorem is immediate: $T_{q,1}=q-1$, the space $V=\F_q$ has $q-1$ nonzero elements,
and each of them recovers every request. We assign distinct servers to the $q-1$ requests. For
$s\ge2$ the proof depends on the parity of $q$, and the two arguments are independent of each
other.

\subsubsection*{Odd \texorpdfstring{$q$}{q}}
By Subsection~\ref{subsec:tworep} it suffices to prove Theorem~\ref{thm:tworep}. For this we
encode the ways of serving a list from the set $S$ of that theorem in the terms of one
determinant. In $S$ we call one of the two chosen vectors of each line $L$ its \emph{row representative}
$\rho_L$ and the other its \emph{column representative} $\kappa_L$. We introduce one indeterminate
$x_L$ for every line $L$ and an $(N+1)\times(N+1)$ matrix $B$ whose rows and columns are indexed
by the lines and one additional index~$0$. The entries of $B$ are linear forms in the $x_L$ with
nonnegative integer coefficients: the coefficient of $x_L$ in the entry of row $I$ and column $J$
is the number of nonzero vectors $w$ of $L$ for which $\rho_I+w$ is a nonzero vector of $J$; in
row~$0$ the zero vector replaces $\rho_I$, and in column~$0$ the condition becomes $\rho_I+w=0$.
A nonzero term in the expansion of $\det B$ chooses a column $\sigma(I)$ and a line $L_I$ for
every row $I$. Such a term yields pairwise disjoint recovery sets: the request on $L_I$ is recovered from
the row representative of $I$ and the column representative of $\sigma(I)$, the index $0$ carries
no server and marks the requests that are recovered from a single server, and because $\sigma$ is a
permutation no representative is used twice. The monomial of the term records how many
requests lie on each line. Consequently a list with $c_L$ requests on each line $L$ is served
by $S$ as soon as the monomial $\prod_Lx_L^{c_L}$ has a nonzero coefficient in $\det B$
(Lemma~\ref{lem:terms}).

Terms of $\det B$ can cancel. It remains to show that every monomial of degree $N+1$
nevertheless occurs in $\det B$; we use the determinant because it can be computed from
eigenvalues. The matrix $B$
represents a convolution operator on the functions on $V$ that are invariant under multiplication
by scalars, and the characters of $V$ diagonalize this operator. This gives a factorization of
$\det B$ into $N+1$ linear forms (Lemma~\ref{lem:det}). We then divide each coefficient of this
product, taken without its constant factor $(-1)^N(q-1)$, by a multinomial coefficient and show
that the ratio, a rational number whose denominator is prime to the characteristic $p$ of
$\F_q$, is congruent to $1$ modulo $p$, so that it is not zero (Lemma~\ref{lem:Dcoeff} and the proof of Proposition~\ref{prop:coeff}). The last step of
this computation
divides by $N+1$, which is congruent to $2$ modulo $p$, and this is where $p$ must be odd.

\subsubsection*{Even \texorpdfstring{$q$}{q}}
For even $q$ neither step of the argument for odd $q$ is available: the number $q-1$ of servers
on a line is odd, so the nonzero vectors of a line cannot be split into pairs as in
Subsection~\ref{subsec:tworep}, and $N+1$ is even, so the division by $N+1$ in the proof of
Proposition~\ref{prop:coeff} fails (see also Section~\ref{sec:conclusion}). The proof for even
$q$ uses two inputs. The first is the theorem of Hall~\cite{Hall}
(Theorem~\ref{thm:hall}): if $z_1,\dots,z_h$ are elements of an abelian group of order $h$ with
sum zero, then the elements of the group have an ordering $x_1,\dots,x_h$ such that
$x_1+z_1,\dots,x_h+z_h$ is again an ordering of all elements. The second input is
Theorem~\ref{thm:bgs}. For $q=2$, Theorem~\ref{thm:main}\ref{main:a} is a short consequence of
Theorem~\ref{thm:bgs} (Lemma~\ref{lem:binary}): a perfect matching with prescribed differences
yields pairwise disjoint recovery sets once the zero vector is removed from the edge that
contains it, and a list whose sum is not zero is reduced to a list with sum zero by changing one
request; the solution is then translated so that this request is recovered from a single server.

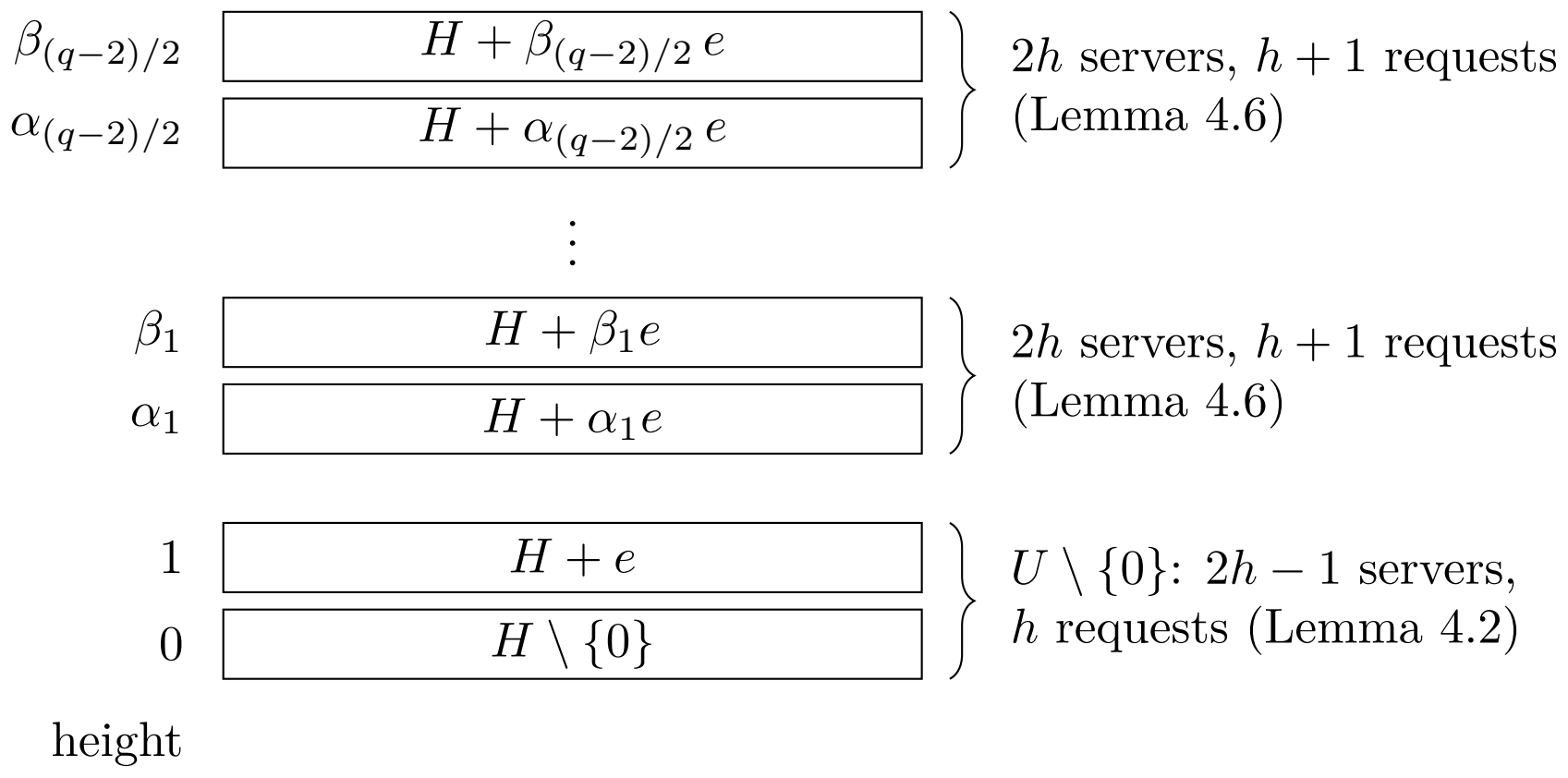
\begin{figure}[ht]
\centering
\begin{tikzpicture}[x=1cm,y=0.62cm,font=\small]
  \foreach \y/\lab in {0/{$0$},1/{$1$},2.6/{$\alpha_1$},3.6/{$\beta_1$},5.9/{$\alpha_{(q-2)/2}$},6.9/{$\beta_{(q-2)/2}$}}
  {
    \draw (0,\y) rectangle (5,\y+0.8);
    \node[anchor=east] at (-0.15,\y+0.4) {\lab};
  }
  \node at (2.5,0.4) {$H\setminus\{0\}$};
  \node at (2.5,1.4) {$H+e$};
  \node at (2.5,3.0) {$H+\alpha_1e$};
  \node at (2.5,4.0) {$H+\beta_1e$};
  \node at (2.5,5.2) {$\vdots$};
  \node at (2.5,6.3) {$H+\alpha_{(q-2)/2}\,e$};
  \node at (2.5,7.3) {$H+\beta_{(q-2)/2}\,e$};
  \node[anchor=east] at (-0.15,-0.75) {height};
  \draw[decorate,decoration={brace,amplitude=5pt,mirror}] (5.2,0) -- (5.2,1.8);
  \node[anchor=west,align=left] at (5.5,0.9) {$U\setminus\{0\}$: $2h-1$ servers,\\ $h$ requests (Lemma~\ref{lem:binary})};
  \draw[decorate,decoration={brace,amplitude=5pt,mirror}] (5.2,2.6) -- (5.2,4.4);
  \node[anchor=west,align=left] at (5.5,3.5) {$2h$ servers, $h+1$ requests\\ (Lemma~\ref{lem:pair})};
  \draw[decorate,decoration={brace,amplitude=5pt,mirror}] (5.2,5.9) -- (5.2,7.7);
  \node[anchor=west,align=left] at (5.5,6.8) {$2h$ servers, $h+1$ requests\\ (Lemma~\ref{lem:pair})};
\end{tikzpicture}
\caption{The servers of $V=\F_q^s$ for even $q\ge4$, arranged in the layers $H+\alpha e$ of a
hyperplane $H$ with $h$ elements. The layers of heights $0$ and $1$ form the group $U$, whose
nonzero elements serve $h$ requests. The other $q-2$ layers are grouped into $(q-2)/2$ pairs of
layers, with heights $\alpha_j$ and $\beta_j$, and the $2h$ servers of each pair serve $h+1$
requests of height~$1$.}
\label{fig:layers}
\end{figure}

For even $q\ge4$ and $s\ge2$ we decompose $V$ into layers, shown in Figure~\ref{fig:layers}
and defined in Subsection~\ref{subsec:layers}. We choose a hyperplane $H$ of $V$ (a subspace of
dimension $s-1$) that contains fewer than half as many requests as it has elements, and a vector
$e\notin H$; such a hyperplane exists by averaging over all hyperplanes
(Lemma~\ref{lem:hyperplane}). Then $V$ is the disjoint union of the $q$ layers $H+\alpha e$,
$\alpha\in\F_q$, and $\alpha$ is the height of the layer. Every request outside $H$ has a multiple
in the layer of height~$1$. The layers of heights $0$ and $1$ together form a subgroup $U$ of $V$
of order $2h$, where $h$ is the number of elements of $H$, and the sum
of two elements of $U$ is a linear combination over $\F_q$ with coefficients equal to~$1$. The
group $U$ is an elementary abelian $2$-group, and Lemma~\ref{lem:binary}, which
also gives the case $q=2$, shows that any $h$ requests in $U$ are served by the nonzero elements
of $U$. The remaining $q-2$ layers are grouped in any way into $(q-2)/2$ pairs. We show that the
$2h$ servers of two layers serve $h+1$ requests $a+e$ of height $1$, two of them from single
servers, provided that the vectors $a\in H$ form what we call a two-layer family
(Lemma~\ref{lem:pair}); this step uses Hall's theorem for the group $H$ and the multiplication of
$\F_q$. Since $H$ contains fewer than $h/2$ requests, enough requests of height $1$ remain for a
counting argument that extracts two-layer families from them for as long as a pair of layers is
unused (Lemma~\ref{lem:extract}); the requests in $H$ and the requests of height $1$ that are
left, $h$ in all, are then served in $U$. In total
\[
  \frac{q-2}{2}\,(h+1)+h=\frac{qh+q-2}{2}=\Tqs
\]
requests are served. In this argument Theorem~\ref{thm:bgs} is applied once, to the group $U$,
and Hall's theorem is applied once for each pair of layers.

\FloatBarrier
\section{Odd characteristic}\label{sec:odd}

In this section $q=p^{\ell}$ is a power of an odd prime $p$, and we prove Theorem~\ref{thm:tworep}. We
keep the notation of Section~\ref{sec:model}. As in Subsection~\ref{subsec:outline}, for every
line $L\in\cP$ we denote the two distinct nonzero vectors of $L$ chosen in
Theorem~\ref{thm:tworep} by $\rho_L$ and $\kappa_L$, the row representative and the column
representative of $L$. Thus $S=\{\rho_L,\kappa_L: L\in\cP\}$.

\subsection{The quotient matrix and the terms of its determinant}\label{subsec:terms}

We define the matrix $B$ of the outline and show that every nonzero term in the expansion of
its determinant yields pairwise disjoint recovery sets for a list of requests. The multiplicative
group $\F_q^{\times}$ acts on $V$ by scalar multiplication. Its orbits are
$\{0\}$ and the sets $L\setminus\{0\}$ with $L\in\cP$. We index the orbits by the set
$\cO\coloneqq\{0\}\cup\cP$, writing $O_0\coloneqq\{0\}$ and $O_L\coloneqq L\setminus\{0\}$, and we
fix the point $u_0\coloneqq0$ of $O_0$ and the point $u_L\coloneqq\rho_L$ of $O_L$. Let $x_L$,
$L\in\cP$, be independent indeterminates. For $I,J\in\cO$ and $L\in\cP$ put
\[
  n^{L}_{IJ}\coloneqq\bigl|\{\,w\in L\setminus\{0\}: u_I+w\in O_J\,\}\bigr|,
  \qquad
  B_{IJ}\coloneqq\sum_{L\in\cP}n^{L}_{IJ}\,x_L ,
\]
and let $B\coloneqq(B_{IJ})_{I,J\in\cO}$. This is a square matrix of size $N+1$ whose entries are
linear forms in the $x_L$ with nonnegative integer coefficients. It is the \emph{quotient
matrix}, in the sense of~\cite[Section~2.3]{BH}, of the convolution operator
$f\mapsto\sum_{L\in\cP}x_L\sum_{w\in L\setminus\{0\}}f(\,\cdot+w)$ with respect to the partition
of $V$ into orbits: it describes the action of this operator on the functions that are constant
on the orbits (proof of Lemma~\ref{lem:det}). Replacing $u_I$ by $\beta u_I$ and $w$ by
$\beta w$, with $\beta\in\F_q^{\times}$, does not change $n^{L}_{IJ}$; hence $B$ does not depend
on the choice of $S$, which enters only through Lemma~\ref{lem:terms}. Since $u_0+w=w\ne0$ for $w\in L\setminus\{0\}$, we
have $B_{00}=0$.

For $c=(c_L)_{L\in\cP}$ with nonnegative integer entries we write
$x^c\coloneqq\prod_{L\in\cP}x_L^{c_L}$ and $|c|\coloneqq\sum_{L\in\cP}c_L$. The determinant of $B$
is a homogeneous polynomial of degree $N+1$ in the $x_L$. The next lemma translates its monomials
into recovery sets.

\begin{lemma}\label{lem:terms}
Let $c=(c_L)_{L\in\cP}$ be nonnegative integers with $|c|=N+1$, and suppose that the monomial
$x^c$ has a nonzero coefficient in $\det B$. Then $S$ serves with locality two every list of
$N+1$ requests that has exactly $c_L$ requests on each line $L$.
\end{lemma}

\begin{proof}
Expand $\det B=\sum_{\sigma}\sgn(\sigma)\prod_{I\in\cO}B_{I\sigma(I)}$ over the permutations
$\sigma$ of $\cO$, and expand every factor $B_{I\sigma(I)}$ into its terms $n^{L}_{I\sigma(I)}x_L$.
Since $x^c$ has a nonzero coefficient, there are a permutation $\sigma$ of $\cO$ and lines $L_I$,
$I\in\cO$, such that $n^{L_I}_{I\sigma(I)}\ge1$ for all $I$ and $\prod_{I\in\cO}x_{L_I}=x^c$. By
the definition of $n^{L}_{IJ}$, for every $I\in\cO$ there is a vector $w_I\in L_I\setminus\{0\}$
with $u_I+w_I\in O_{\sigma(I)}$. We attach a set $R_I\subseteq S$ to every $I\in\cO$.
\begin{enumerate}[label=\textup{(\roman*)}]
\item If $I\ne0$ and $J\coloneqq\sigma(I)\ne0$, then $\rho_I+w_I$ is a nonzero vector of $J$, so
$\rho_I+w_I=\beta\kappa_J$ for some $\beta\in\F_q^{\times}$, and $w_I=\beta\kappa_J-\rho_I$. We
put $R_I\coloneqq\{\rho_I,\kappa_J\}$. The two servers are different: they lie on different lines
if $I\ne J$, and $\rho_I\ne\kappa_I$ by the choice of the representatives if $I=J$.
\item If $I=0$, then $w_0=u_0+w_0\in O_{\sigma(0)}$ and $w_0\ne0$, so $J\coloneqq\sigma(0)\ne0$
and $w_0=\beta\kappa_J$ for some $\beta\in\F_q^{\times}$. We put $R_0\coloneqq\{\kappa_J\}$.
\item If $I\ne0$ and $\sigma(I)=0$, then $\rho_I+w_I=0$, so $w_I=-\rho_I$. We put
$R_I\coloneqq\{\rho_I\}$.
\end{enumerate}
In each case $w_I$ is a linear combination with nonzero coefficients of the elements of $R_I$.
The sets $R_I$ are pairwise disjoint, because $\rho_I$ belongs only to $R_I$ and $\kappa_J$
belongs only to $R_{\sigma^{-1}(J)}$.

Now consider a list with $c_L$ requests on each line $L$. Exactly $c_L$ of the indices $I\in\cO$
satisfy $L_I=L$, so we can assign the requests of the list bijectively to the indices in such a
way that a request assigned to $I$ lies on $L_I$. Such a request is a nonzero multiple of $w_I$
and is therefore a linear combination with nonzero coefficients of the elements of $R_I$.
\end{proof}

\begin{example}\label{ex:plane}
To illustrate $B$ and Lemma~\ref{lem:terms}, let $s=2$, so that $N=q+1$. For two different lines $I,J$ and a line $L\notin\{I,J\}$, the affine
line $\rho_I+L$ meets $J$ in exactly one point, and this point is nonzero; for $L\in\{I,J\}$ the
set $\rho_I+(L\setminus\{0\})$ contains no nonzero vector of $J$. Writing $X$ for the sum of all
$x_L$, we obtain
\begin{gather*}
  B_{00}=0,\qquad B_{0J}=(q-1)\,x_J,\qquad B_{I0}=x_I,\\
  B_{II}=(q-2)\,x_I,\qquad B_{IJ}=X-x_I-x_J\quad(I\ne J)
\end{gather*}
for lines $I,J$. Take $q=3$, so that there are four lines $L_1,\dots,L_4$, and consider five
requests on $L_1$, that is, the monomial $x_{L_1}^5$. The entries of $B$ that contain $x_{L_1}$
are $B_{0L_1}$, $B_{L_10}$, $B_{L_1L_1}$ and $B_{IJ}$ for different $I,J\in\{L_2,L_3,L_4\}$. A
term of the expansion that produces $x_{L_1}^5$ is given by $\sigma(0)=L_1$, $\sigma(L_1)=0$,
$\sigma(L_2)=L_3$, $\sigma(L_3)=L_4$, $\sigma(L_4)=L_2$. By the proof of Lemma~\ref{lem:terms}
this term gives the following recovery sets for the five requests:
\[
  \{\kappa_{L_1}\},\quad\{\rho_{L_1}\},\quad\{\rho_{L_2},\kappa_{L_3}\},\quad
  \{\rho_{L_3},\kappa_{L_4}\},\quad\{\rho_{L_4},\kappa_{L_2}\}.
\]
Two requests are recovered from the two servers on $L_1$, and the six servers on the other lines
form three pairs, in agreement with the count in Proposition~\ref{prop:upper}.
\end{example}

\subsection{The determinant of the quotient matrix}\label{subsec:det}

By Lemma~\ref{lem:terms} it remains to determine which monomials of degree $N+1$ occur in
$\det B$. In this subsection we factor $\det B$ into linear forms. Let $V^{*}$ be the space of linear functionals $\xi\colon V\to\F_q$. Two nonzero functionals have
the same kernel if and only if they are proportional, so the $N$ one-dimensional subspaces
$[\xi]$ of $V^{*}$ correspond to the $N$ hyperplanes $\ker\xi$ of $V$. Put
\[
  X\coloneqq\sum_{L\in\cP}x_L ,\qquad
  X_{\xi}\coloneqq\sum_{L\in\cP,\ L\subseteq\ker\xi}x_L\quad(\xi\in V^{*}\setminus\{0\}).
\]
The linear form $X_{\xi}$ depends only on $[\xi]$.

\begin{lemma}\label{lem:det}
We have
\[
  \det B=(-1)^N\,(q-1)\,X\,D,\qquad\text{where}\quad
  D\coloneqq\prod_{[\xi]}\bigl(X-q\,X_{\xi}\bigr)
\]
and the product is taken over the $N$ one-dimensional subspaces $[\xi]$ of $V^{*}$.
\end{lemma}

\begin{proof}
Both sides are polynomials in the $x_L$ with integer coefficients, so it suffices to prove the
identity after the $x_L$ have been replaced by arbitrary complex numbers. Fix such numbers, and
let $A$ be the convolution operator on the space of functions $f\colon V\to\C$ defined by
\[
  (Af)(u)\coloneqq\sum_{L\in\cP}x_L\sum_{w\in L\setminus\{0\}}f(u+w).
\]

\emph{The matrix $B$ represents $A$ on the invariant functions.} Call $f$ \emph{invariant} if
$f(\beta u)=f(u)$ for all $u\in V$ and $\beta\in\F_q^{\times}$. The invariant functions form
a space $\cW$ of dimension $N+1$, and the indicator functions $1_J$ of the orbits $O_J$,
$J\in\cO$, are a basis of it. If $f$ is invariant, then so is $Af$: substituting
$w=\beta w'$, where $w'$ runs through $L\setminus\{0\}$ together with $w$, we get
\[
  (Af)(\beta u)=\sum_{L}x_L\sum_{w'\in L\setminus\{0\}}f\bigl(\beta(u+w')\bigr)=(Af)(u).
\]
Moreover $(A1_J)(u_I)=\sum_{L}x_L\,|\{w\in L\setminus\{0\}:u_I+w\in O_J\}|=B_{IJ}$. Since $A1_J$ is
constant on the orbits, this means $A1_J=\sum_{I\in\cO}B_{IJ}1_I$. Hence $B$ is the matrix of the
restriction $A|_{\cW}$ in the basis $(1_J)_{J\in\cO}$, and $\det B=\det(A|_{\cW})$.

\emph{Eigenvectors.} Fix a nontrivial homomorphism $\psi$ from the additive group of $\F_q$ to
$\C^{\times}$. Then $\sum_{a\in\F_q}\psi(a)=0$, because this sum is unchanged when it is
multiplied by $\psi(b)$ for any $b$, and $\psi(b)\ne1$ for some $b$. For $\xi\in V^{*}$ let
$\chi_{\xi}(u)\coloneqq\psi(\xi(u))$. The $q^s$ functions $\chi_{\xi}$ are characters of $V$,
that is, homomorphisms from $V$ to $\C^{\times}$. They are pairwise distinct, because for $\xi\ne\xi'$ the
functional $\xi-\xi'$ maps $V$ onto $\F_q$ and $\psi$ is nontrivial; distinct homomorphisms to
$\C^{\times}$ are linearly independent. Since
$\chi_{\xi}(u+w)=\chi_{\xi}(u)\chi_{\xi}(w)$, we have $A\chi_{\xi}=\lambda_{\xi}\chi_{\xi}$ with
\[
  \lambda_{\xi}\coloneqq\sum_{L\in\cP}x_L\sum_{w\in L\setminus\{0\}}\psi(\xi(w)).
\]
If $L\subseteq\ker\xi$, the inner sum is $q-1$. Otherwise $\xi$ maps $L\setminus\{0\}$
bijectively onto $\F_q^{\times}$, and the inner sum is $\sum_{a\ne0}\psi(a)=-1$. Therefore
$\lambda_0=(q-1)X$ and $\lambda_{\xi}=(q-1)X_{\xi}-(X-X_{\xi})=qX_{\xi}-X$ for $\xi\ne0$.

For $\beta\in\F_q^{\times}$ we have $\chi_{\xi}(\beta u)=\chi_{\beta\xi}(u)$. Hence the
function $\chi_0=1$ and the $N$ functions
$\sum_{\beta\in\F_q^{\times}}\chi_{\beta\xi}$, one for each one-dimensional subspace $[\xi]$ of
$V^{*}$, are invariant. They are sums over pairwise disjoint sets of the functions $\chi_{\xi}$,
so they are linearly independent, and since there are $N+1$ of them they form a basis of $\cW$.
Each of them is an eigenvector of $A$, with eigenvalue $(q-1)X$ and $qX_{\xi}-X$ respectively,
because $X_{\beta\xi}=X_{\xi}$. Consequently
\[
  \det(A|_{\cW})=(q-1)X\prod_{[\xi]}\bigl(qX_{\xi}-X\bigr)=(-1)^N(q-1)\,X\,D. \qedhere
\]
\end{proof}

\subsection{Every monomial of degree \texorpdfstring{$N+1$}{N+1} occurs in \texorpdfstring{$\det B$}{det B}}\label{subsec:coeff}

By Lemmas~\ref{lem:terms} and~\ref{lem:det}, Theorem~\ref{thm:tworep} follows from the next
proposition.

\begin{proposition}\label{prop:coeff}
For all nonnegative integers $c=(c_L)_{L\in\cP}$ with $|c|=N+1$, the coefficient of $x^c$ in the
polynomial $X\,D$ is nonzero.
\end{proposition}

Modulo $q$ every factor $X-qX_{\xi}$ of $D$ is $X$, so the proof compares the coefficients of
$XD$ with those of $X^{N+1}$, which are multinomial coefficients, and works modulo $p$. For a
nonzero rational number $a$ let $v_p(a)$ be its $p$-adic valuation, and let $\Z_{(p)}$ be the ring
of rational numbers whose denominator is not divisible by $p$. For integers $n\ge j\ge0$ let
$(n)_j\coloneqq n(n-1)\cdots(n-j+1)$, with $(n)_0\coloneqq1$. In the proof of
Lemma~\ref{lem:Dcoeff} a coefficient of $D$, divided by a multinomial coefficient, is written as
a sum of contributions; one of them is $1$, and each of the others carries a factor $q^j/(N)_j$
with $1\le j\le N$. The following estimate shows that the $p$-adic valuation of this factor is
positive. It is stated for the
numbers $N_s\coloneqq1+q+\dots+q^{s-1}$; note that $N=N_s$.

\begin{lemma}\label{lem:valuation}
For all $s\ge1$ and $1\le j\le N_s$ we have $v_p\bigl((N_s)_j\bigr)<\ell j$, where $q=p^{\ell}$.
\end{lemma}

\begin{proof}
We use induction on $s$. For $s=1$ we have $N_1=1$ and $(N_1)_1=1$. Let $s>1$, so that
$N_s=1+qN_{s-1}$, and consider the $j$ factors $N_s-i$, $0\le i\le j-1$, of $(N_s)_j$. They are
positive integers, and $N_s-i\equiv1-i\pmod q$. Call the index $i$ \emph{special} if
$i\equiv1\pmod q$. A factor with a special index $i=1+qk$ equals $q(N_{s-1}-k)$. A factor whose
index is not special is not divisible by $q$, so its valuation is at most $\ell-1$. If $i=1+qk$ is
special, then $i-1=qk$ is not special, it belongs to the range of indices, and the factor
$N_s-qk=1+q(N_{s-1}-k)$ has valuation~$0$.

Let $b$ be the number of special indices; they are the numbers $1+qk$ with $0\le k\le b-1$, and
$b\le N_{s-1}$ because $1+qk\le j-1\le qN_{s-1}$. If $b=0$, then all $j$ factors have valuation at
most $\ell-1$, and $v_p((N_s)_j)\le(\ell-1)j<\ell j$. If $b\ge1$, the product of the factors with special
index is $q^b(N_{s-1})_b$, and by the induction hypothesis its valuation is
$\ell b+v_p((N_{s-1})_b)<2\ell b$. The $b$ factors with index $qk$, $0\le k\le b-1$, have
valuation $0$, and each of the remaining $j-2b$ factors has valuation less than $\ell$.
Altogether $v_p((N_s)_j)<2\ell b+\ell(j-2b)=\ell j$.
\end{proof}

For nonnegative integers $a=(a_L)_{L\in\cP}$ we write
$\binom{|a|}{a}\coloneqq|a|!/\prod_{L}a_L!$ for the multinomial coefficient.

\begin{lemma}\label{lem:Dcoeff}
For all nonnegative integers $a=(a_L)_{L\in\cP}$ with $|a|=N$, the coefficient of $x^a$ in $D$ is
$\binom{N}{a}\,\theta_a$ for a rational number $\theta_a$ that belongs to $1+p\,\Z_{(p)}$.
\end{lemma}

\begin{proof}
The polynomial $D$ is a product of $N$ factors $X-qX_{\xi}$. We expand it by choosing in every
factor either $X$ or $-qX_{\xi}$, and then, in every factor where $-qX_{\xi}$ was chosen, one of
the terms $x_L$ of $X_{\xi}$. A choice in which $-qX_{\xi}$ is taken from $j$ factors and the
selected terms are $x_{L_1},\dots,x_{L_j}$ contributes $(-q)^j\,x^bX^{N-j}$, where $b_L$ is the
number of indices $i$ with $L_i=L$. The coefficient of $x^a$ in $x^bX^{N-j}$ is the multinomial
coefficient $(N-j)!/\prod_L(a_L-b_L)!$ if $b_L\le a_L$ for all $L$, and it is $0$ otherwise.
Dividing by $\binom{N}{a}$, we see that the contribution of this choice to $\theta_a$ is $0$ or
\[
  (-q)^j\,\frac{(N-j)!}{N!}\prod_{L}\frac{a_L!}{(a_L-b_L)!}
  =(-q)^j\,\frac{\prod_L(a_L)_{b_L}}{(N)_j}.
\]
For $j=0$ there is one choice, and its contribution is $1$. For $j\ge1$ the numerator
$\prod_L(a_L)_{b_L}$ is an integer and, by Lemma~\ref{lem:valuation},
$v_p\bigl(q^j/(N)_j\bigr)=\ell j-v_p((N)_j)>0$, so the contribution belongs to $p\,\Z_{(p)}$. The sum
of all contributions is $\theta_a$.
\end{proof}

\begin{proof}[Proof of Proposition~\ref{prop:coeff}]
Let $e_L$ denote the vector with entry $1$ at $L$ and $0$ elsewhere. The coefficient of $x^c$ in
$XD=\sum_Lx_LD$ is the sum, over the lines $L$ with $c_L\ge1$, of the coefficient of
$x^{c-e_L}$ in $D$. By Lemma~\ref{lem:Dcoeff} and the identity
$\binom{N}{c-e_L}=\frac{c_L}{N+1}\binom{N+1}{c}$, this coefficient equals
\[
  \binom{N+1}{c}\sum_{L:\,c_L\ge1}\frac{c_L}{N+1}\;\theta_{c-e_L}.
\]
Now $N+1=2+q+\dots+q^{s-1}\equiv2\pmod p$, and $p$ is odd, so $N+1$ is invertible in $\Z_{(p)}$
and the weights $c_L/(N+1)$ belong to $\Z_{(p)}$. Their sum is $|c|/(N+1)=1$, and every
$\theta_{c-e_L}$ belongs to $1+p\,\Z_{(p)}$. Hence the sum belongs to $1+p\,\Z_{(p)}$; in
particular it is not zero, and neither is the coefficient.
\end{proof}

\begin{proof}[Proof of Theorem~\ref{thm:tworep}]
Let a list of $N+1$ requests be given, and let $c_L$ be the number of its requests on the line
$L$. By Lemma~\ref{lem:det} and Proposition~\ref{prop:coeff} the monomial $x^c$ has a nonzero
coefficient in $\det B$, and Lemma~\ref{lem:terms} shows that $S$ serves the list with locality
two.
\end{proof}

Together with the deductions in Subsections~\ref{subsec:tworep} and~\ref{subsec:length}, this
completes the proof of Theorem~\ref{thm:main}\ref{main:a} and of
Corollary~\ref{cor:length}\ref{length:a} for odd $q$, and the proof of
Corollary~\ref{cor:length}\ref{length:b}.

\section{Characteristic two}\label{sec:even}

In this section $q$ is a power of two, and we prove Theorem~\ref{thm:main}\ref{main:a} for these
$q$. In characteristic two every element is its own negative, so sums and differences coincide.

\subsection{Hall's theorem and the case \texorpdfstring{$q=2$}{q=2}}\label{subsec:inputs}

We first state Hall's theorem in the form in which it is used.

\begin{theorem}[Hall~\cite{Hall}; see {\cite[Theorem~4.4]{HKRS}}]\label{thm:hall}
Let $A$ be a finite abelian group of order $h$, and let $z_1,\dots,z_h$ be elements of $A$ with
$z_1+\dots+z_h=0$. Then there is an ordering $x_1,\dots,x_h$ of the elements of $A$ such that
$x_1+z_1,\dots,x_h+z_h$ is again an ordering of the elements of $A$.
\end{theorem}

Here an ordering of the elements of $A$ lists every element exactly once. The elements $z_i$
need not be distinct and may be zero. We apply the theorem only to elementary abelian
$2$-groups, that is, abelian groups in which every element has order at most two; such a group is
the additive group of a vector space over $\F_2$. In such a group the elements
$y_i\coloneqq x_i+z_i$ satisfy $x_i+y_i=z_i$, so the theorem provides two orderings
$x_1,\dots,x_h$ and $y_1,\dots,y_h$ of $A$ with $x_i+y_i=z_i$ for all $i$.

The second input of the outline is Theorem~\ref{thm:bgs}. We use it through the following
consequence, in which
an arbitrary list replaces the list with sum zero.

\begin{lemma}\label{lem:binary}
Let $U$ be an elementary abelian $2$-group of order $2h\ge4$, and let $r_1,\dots,r_h$ be nonzero
elements of $U$. Then there are pairwise disjoint subsets $R_1,\dots,R_h$ of $U\setminus\{0\}$,
each of size one or two, such that $r_i$ is the sum of the elements of $R_i$ for every $i$.
\end{lemma}

\begin{proof}
We identify $U$ with $\F_2^n$, where $2^n=2h$ and $n\ge2$. Let $\Sigma\coloneqq r_1+\dots+r_h$.

Suppose first that $\Sigma=0$. By Theorem~\ref{thm:bgs} there are $x_1,\dots,x_h$ such that the
pairs $\{x_i,x_i+r_i\}$ partition $U$. Exactly one of these pairs contains $0$, say the pair with
index $k$; then this pair is $\{0,r_k\}$. Put $R_k\coloneqq\{r_k\}$ and
$R_i\coloneqq\{x_i,x_i+r_i\}$ for $i\ne k$.

Suppose now that $\Sigma\ne0$. Some $r_i$ is different from $\Sigma$: otherwise
$\Sigma=h\Sigma=0$, because $h$ is even. Let $r_j\ne\Sigma$, and replace $r_j$ by
$r_j'\coloneqq r_j+\Sigma$, which is nonzero. The new list has sum $\Sigma+\Sigma=0$, so by
Theorem~\ref{thm:bgs} there is a partition of $U$ into pairs $\{x_i,x_i+r_i\}$, $i\ne j$, and
$\{x_j,x_j+r_j'\}$. Adding the element $x_j+r_j$ to all members of all pairs gives another
partition of $U$ into pairs with the same sums, in which the pair with index $j$ is
$\{r_j,\Sigma\}$. We may therefore assume that $x_j=r_j$ and $x_j+r_j'=\Sigma$. The element $0$
belongs to a pair with an index $k\ne j$, and this pair is $\{0,r_k\}$. Put $R_j\coloneqq\{r_j\}$,
$R_k\coloneqq\{r_k\}$ and $R_i\coloneqq\{x_i,x_i+r_i\}$ for $i\notin\{j,k\}$. These sets are
pairwise disjoint subsets of $U\setminus\{0\}$; the element $\Sigma$ belongs to none of them.
\end{proof}

A similar reduction, in which one request is changed so that the sum becomes zero and the
solution is then translated, appears in the proof of~\cite[Lemma~4.10]{KRS}.

\begin{proof}[Proof of Theorem~\ref{thm:main}\ref{main:a} for $q=2$]
For $s=1$ see Subsection~\ref{subsec:outline}. For $s\ge2$ apply Lemma~\ref{lem:binary} to
$U=\F_2^s$, where $h=2^{s-1}=T_{2,s}$.
\end{proof}

\begin{remark}\label{rem:hall}
For an elementary abelian $2$-group $A$ of order $h\ge2$, Theorem~\ref{thm:hall} is also a
special case of Theorem~\ref{thm:bgs}. Indeed, the $h$ elements $(z_i,1)$ of $A\times\F_2$ are
nonzero and have sum zero, and in a partition of $A\times\F_2$ into pairs with these sums every
pair consists of an element $(x_i,0)$ and the element $(x_i+z_i,1)$, so that both the $x_i$ and
the $x_i+z_i$ run through $A$. We quote
Hall's theorem separately because it is classical and because the argument of
Subsection~\ref{subsec:twolayers} needs nothing more.
\end{remark}

\subsection{Layers}\label{subsec:layers}

From now on let $q\ge4$ be a power of two and $s\ge2$; the case $s=1$ is treated in
Subsection~\ref{subsec:outline}. Put $h\coloneqq q^{s-1}$, a multiple of $4$, so that
$\Tqs=(qh+q-2)/2$. Every hyperplane of $V$, that is, every subspace of dimension $s-1$, has $h$
elements. We build the layers on a hyperplane that contains few requests, so that enough requests
lie outside it for all pairs of layers (Subsection~\ref{subsec:assembly}); the next lemma provides
such a hyperplane.

\begin{lemma}\label{lem:hyperplane}
For every list of $\Tqs$ requests there is a hyperplane of $V$ that contains at most $h/2-1$
requests of the list.
\end{lemma}

\begin{proof}
There are $N=(qh-1)/(q-1)$ hyperplanes, and a nonzero vector lies in $(h-1)/(q-1)$ of them.
Hence the average number of requests in a hyperplane is $\Tqs(h-1)/(qh-1)$. This is smaller than
$h/2$, because
\[
  h(qh-1)-2\Tqs(h-1)=h(qh-1)-(qh+q-2)(h-1)=h+q-2>0 .
\]
Some hyperplane therefore contains fewer than $h/2$ requests, that is, at most $h/2-1$ of them.
\end{proof}

Fix a list of $\Tqs$ requests, a hyperplane $H$ as in Lemma~\ref{lem:hyperplane}, and a vector
$e\in V\setminus H$. Let $n_H\le h/2-1$ be the number of requests in $H$. Every vector of $V$ has
a unique expression $w+\alpha e$ with $w\in H$ and $\alpha\in\F_q$. We call $\alpha$ its
\emph{height}, and we call $H_{\alpha}\coloneqq H+\alpha e$ the \emph{layer} of height $\alpha$
(see Figure~\ref{fig:layers}).
The servers are the elements of $H_0\setminus\{0\}$ and of the layers $H_{\alpha}$ with
$\alpha\ne0$. A request $r$ outside $H$ has a nonzero height $\alpha$, and its multiple
$\alpha^{-1}r$ has height $1$. By the remarks after Definition~\ref{def:serve} we may replace $r$
by this multiple. After this replacement the list consists of $n_H$ requests in $H$ and of
$\Tqs-n_H$ requests of the form $a+e$ with $a\in H$. We call $a$ the \emph{point} of the request
$a+e$.

The set $U\coloneqq H_0\cup H_1$ is closed under addition, since $(w+e)+(w'+e)=w+w'$. It is an
elementary abelian $2$-group of order $2h$, and a sum of two of its elements is a linear
combination over $\F_q$ with both coefficients equal to $1$. The $q-2$ heights different from $0$
and $1$ are grouped into $(q-2)/2$ pairs $\{\alpha_j,\beta_j\}$, which is possible because $q$ is
even; we call $H_{\alpha_j},H_{\beta_j}$ a pair of layers.

\subsection{Two-layer families}\label{subsec:twolayers}

We now describe the families of points $a$ for which the two layers of a pair serve the requests
$a+e$. Two layers contain $2h$ servers. We shall recover two requests from single servers and
$h-1$ requests from two-element sets that consist of one server of each layer. These sets come
from $h-1$ of the $h$ pairs $(x_i,y_i)$ that Hall's theorem provides for $h$ elements $z_i$ of
$H$ with sum zero. The remaining pair gives the two single servers. The servers built from $x_i$
and $y_i$ are multiples of $x_i/c+e$ and of $y_i/(1+c)+e$, for a fixed $c\in\F_q\setminus\{0,1\}$;
since $c+(1+c)=1$, the sum of $c$ times the first and $1+c$ times the second is
$x_i+y_i+e=z_i+e$. The remaining pair recovers two requests with points $a$ and $b$ when it is
$(ca,(1+c)b)$. Its
sum is then $d=ca+(1+c)b$, a third point of the affine line through $a$ and $b$, or $d=a$ if
$a=b$. A third request with the point $d$ balances this sum, so the
points of the other $h-2$ requests must have sum zero. Definition~\ref{def:twolayer} records
these conditions. We use families of points indexed by a
finite set, so that a point may occur several times, and a subfamily is the restriction to a
subset of the index set. Three points $a,b,d$ of $H$ are
\emph{collinear} if they lie on a common affine line of the $\F_q$-space $H$; for three distinct
points this means $d=ca+(1+c)b$ for some $c\in\F_q\setminus\{0,1\}$. If exactly two of the three
points are equal, no such $c$ exists, and Definition~\ref{def:twolayer} excludes this case.

\begin{definition}\label{def:twolayer}
A family of $h+1$ points of $H$ is a \emph{two-layer family} if it consists of three points
$a,b,d$ that are either all equal or pairwise distinct and collinear, and of $h-2$ further points
whose sum is zero.
\end{definition}

The name is explained by the next lemma: the requests whose points form such a family are served
by two layers, and they use all servers of these layers.

\begin{lemma}\label{lem:pair}
Let $\alpha$ and $\beta$ be two different nonzero elements of $\F_q$, and let $(a_i)_{i\in Q}$ be
a two-layer family of points of $H$. Then the requests $a_i+e$, $i\in Q$, are served with
locality two by the $2h$ servers of $H_{\alpha}\cup H_{\beta}$.
\end{lemma}

\begin{proof}
Let $a,b,d$ be the three points and $z_3,\dots,z_h$ the further $h-2$ points of the family, so
that $z_3+\dots+z_h=0$. Choose $c\in\F_q\setminus\{0,1\}$ with $d=ca+(1+c)b$. If the three points
are pairwise distinct and collinear, such a $c$ exists, as noted before
Definition~\ref{def:twolayer}; if they are all equal, every $c\in\F_q\setminus\{0,1\}$ has this
property, and such elements exist because $q\ge4$.

Put $z_1\coloneqq d$ and $z_2\coloneqq d$. The element $z_2$ stands for the request $d+e$, and
$z_1$ stands for the two requests $a+e$ and $b+e$ together. Since $z_1+z_2=0$, the $h$ elements
$z_1,\dots,z_h$ of $H$ have sum zero. By Theorem~\ref{thm:hall} and the paragraph following it,
applied to the additive group of $H$, there are orderings $x_1,\dots,x_h$ and $y_1,\dots,y_h$
of $H$ with $x_i+y_i=z_i$ for all $i$. Adding a fixed element of $H$ to all $x_i$ and all $y_i$
gives two orderings with the same property, so we may assume that $x_1=ca$. Then
$y_1=d+ca=(1+c)b$.

For $1\le i\le h$ consider the servers
\[
  \sigma_i\coloneqq\frac{\alpha}{c}\,x_i+\alpha e\in H_{\alpha},\qquad
  \sigma_i'\coloneqq\frac{\beta}{1+c}\,y_i+\beta e\in H_{\beta}.
\]
The maps $x\mapsto(\alpha/c)x$ and $y\mapsto(\beta/(1+c))y$ are bijections of $H$. Hence
$\sigma_1,\dots,\sigma_h$ are the elements of $H_{\alpha}$, each listed once, and
$\sigma_1',\dots,\sigma_h'$ are the elements of $H_{\beta}$, each listed once; the two layers are
disjoint because $\alpha\ne\beta$. For $2\le i\le h$ we have
\[
  \frac{c}{\alpha}\,\sigma_i+\frac{1+c}{\beta}\,\sigma_i'
  =(x_i+ce)+(y_i+(1+c)e)=z_i+e ,
\]
so the request $z_i+e$ is a linear combination with nonzero coefficients of $\sigma_i$ and
$\sigma_i'$. For $i=1$ we have $\sigma_1=\alpha(a+e)$ and $\sigma_1'=\beta(b+e)$, so the request
$a+e$ is recovered from $\{\sigma_1\}$ and the request $b+e$ from $\{\sigma_1'\}$. The recovery
sets $\{\sigma_1\}$, $\{\sigma_1'\}$ and $\{\sigma_i,\sigma_i'\}$, $2\le i\le h$, are pairwise
disjoint, and they serve the $h+1$ requests $a+e$, $b+e$, $d+e$, $z_3+e,\dots,z_h+e$. If $a=b$,
the two requests $a+e$ and $b+e$ are equal, and they are recovered from the two different
servers $\sigma_1$ and $\sigma_1'$.
\end{proof}

A construction of this kind appears in~\cite[Theorem~3.1]{HKRS}: it serves the lists of $2^{s-1}$
requests that avoid a fixed hyperplane of $\F_2^s$, and each request is recovered from a server
outside the hyperplane together with at most one server inside it.

To apply Lemma~\ref{lem:pair} to every pair of layers we must find, among the requests of
height $1$ that have not yet been served, $h+1$ requests whose points form a two-layer family.
The next lemma shows that this is possible whenever at least $3h/2+2$ such requests remain, and
the bound $n_H\le h/2-1$ of Lemma~\ref{lem:hyperplane} guarantees this number until the last pair
of layers has been treated (Subsection~\ref{subsec:assembly}).

\begin{lemma}\label{lem:extract}
Every family of at least $3h/2+2$ points of $H$ has a subfamily that is a two-layer family.
\end{lemma}

\begin{proof}
Let $(a_i)_{i\in Q}$ be the family, $|Q|\ge3h/2+2$. We select $h+1$ indices in three steps.

\emph{A triple.} If some point occurs at least three times, we select three indices with this
point. Otherwise every point occurs at most twice, so the family contains at least
$(3h/2+2)/2>h/2$ distinct points. The $h$ points of $H$ are partitioned into $h/q$ parallel affine
lines, and $h/2\ge2h/q$ because $q\ge4$. Hence one of these lines contains three distinct points
of the family, and we select three indices with these points.

\emph{A pair.} After the triple has been removed, at least $3h/2-1>h$ indices remain, where we
use $h\ge4$. Two of them carry the same point. We select them; the sum of their points is zero.

\emph{Quadruples.} It remains to select $h-4$ further indices whose points have sum zero. Since
$4$ divides $h$, it suffices to select $(h-4)/4$ times in succession four indices whose points
have sum zero. Before each of these selections at most $5+(h-8)$ indices have been removed, so at
least $(3h/2+2)-5-(h-8)=h/2+5$ indices are available. We claim that a family
$(a_i)_{i\in Q'}$ of points of $H$ in which no four distinct indices have points with sum zero
satisfies $|Q'|<h/2+5$; this completes the selection.

To prove the claim, let $g$ be the number of distinct points of such a family. No point occurs
four times, and at most one point occurs more than once, because two equal points together with
two other equal points have sum zero. Hence $|Q'|\le g+2$. For two different pairs $\{a,b\}$ and
$\{a',b'\}$ of distinct points of the family we have $a+b\ne a'+b'$: if the pairs have a common
point, the equation forces them to coincide, and if they are disjoint, the equation says that four
distinct points have sum zero. The sums $a+b$ are nonzero elements of $H$, so $\binom g2\le h-1$,
which gives $g\le(1+\sqrt{8h-7})/2$. Finally $(1+\sqrt{8h-7})/2+2<h/2+5$, because this
inequality is equivalent to $\sqrt{8h-7}<h+5$, and $8h-7<(h+5)^2$.

The selected indices are distinct, and their number is $3+2+(h-4)=h+1$. The points of the pair
and of the quadruples form $h-2$ points with sum zero.
\end{proof}

\subsection{Combining the layers}\label{subsec:assembly}

The proof of Theorem~\ref{thm:main}\ref{main:a} for even $q\ge4$ combines
Lemmas~\ref{lem:hyperplane}, \ref{lem:pair}, \ref{lem:extract} and~\ref{lem:binary}.

\begin{proof}[Proof of Theorem~\ref{thm:main}\ref{main:a} for even $q\ge4$ and $s\ge2$]
We use the notation of Subsection~\ref{subsec:layers}. After the replacement described there, the
list consists of $n_H\le h/2-1$ requests in $H$ and of the requests $a_i+e$, $i\in Q$, where
$|Q|=\Tqs-n_H$. Recall that $\Tqs=\frac{q-2}{2}(h+1)+h$.

We treat the pairs of layers $H_{\alpha_j},H_{\beta_j}$, $1\le j\le(q-2)/2$, one after the other.
When the pair with index $j$ is treated, $(j-1)(h+1)$ indices of $Q$ have been removed, so the
number of remaining indices is at least
\[
  \Tqs-n_H-\Bigl(\frac{q-2}{2}-1\Bigr)(h+1)=2h+1-n_H\ \ge\ \frac{3h}{2}+2 .
\]
By Lemma~\ref{lem:extract} the remaining points have a subfamily that is a two-layer family. By
Lemma~\ref{lem:pair} the corresponding $h+1$ requests are served with locality two by the servers
of $H_{\alpha_j}\cup H_{\beta_j}$. We remove their indices from $Q$.

After all pairs of layers have been treated, $\Tqs-n_H-\frac{q-2}{2}(h+1)=h-n_H$ requests of
height $1$ remain. Together with the $n_H$ requests in $H$ they are $h$ nonzero elements of the
group $U=H_0\cup H_1$. By Lemma~\ref{lem:binary} they are served by pairwise disjoint subsets of
$U\setminus\{0\}$ of size one or two, each request being the sum of the elements of its set.
These servers lie in the layers of heights $0$ and $1$, which were not used before, and the pairs
of layers used before are disjoint from each other. All $\Tqs$ requests are therefore served with
locality two. Each original request is a nonzero multiple of the request that replaced it, so it
is a linear combination with nonzero coefficients of the same servers.
\end{proof}

This completes the proof of Theorem~\ref{thm:main}\ref{main:a}, and with it the proof of
Corollary~\ref{cor:length}\ref{length:a}. The case $s=1$ is treated in
Subsection~\ref{subsec:outline}, odd $q$ in Subsection~\ref{subsec:tworep} and
Section~\ref{sec:odd}, the case $q=2$ in Subsection~\ref{subsec:inputs}, and even $q\ge4$ in the
present subsection. The last two cases depend on Theorem~\ref{thm:bgs}, which is proved in
Appendix~\ref{app:bgs}.

\section{Concluding remarks}\label{sec:conclusion}

We have shown that the nonzero vectors of $\F_q^s$ serve every list of $(q^s+q-2)/2$ requests
with recovery sets of size at most two, for every finite field and every $s\ge1$, and that this
list size is the largest possible. The proofs for odd and for even $q$ are independent of each
other. For odd
$q$ the argument uses a determinant whose entries are linear forms in one variable for each
line: every nonzero term in its expansion yields pairwise disjoint recovery sets, every monomial
of degree $N+1$ occurs in it with a nonzero coefficient, and this gives the stronger
Theorem~\ref{thm:tworep} on two servers per line. For even $q$ the statement
is reduced to Theorem~\ref{thm:bgs} on perfect matchings of $\F_2^n$ with prescribed
differences, directly for $q=2$ and, for $q\ge4$, through the layers of a hyperplane and the
theorem of M.~Hall~\cite{Hall}; Theorem~\ref{thm:bgs} is proved in~\cite{Bin} by determining the $2$-adic valuation of a
signed count of such matchings with labelled pairs, and the proof in Appendix~\ref{app:bgs}
follows~\cite{Bin}.

Two questions arise directly from these results.

\emph{Finding the recovery sets.} The proof of Theorem~\ref{thm:tworep} and the proof of
Theorem~\ref{thm:bgs} in Appendix~\ref{app:bgs} are existential: they show
that certain coefficients are nonzero. The other steps of the proof for even $q$ are constructive: the hyperplane of
Lemma~\ref{lem:hyperplane} is found by inspecting all hyperplanes, the theorem of M.~Hall has an
algorithmic proof~\cite{HKRS}, the proofs of Lemmas~\ref{lem:pair} and~\ref{lem:extract} are
explicit, and so is the reduction of Lemma~\ref{lem:binary} to Theorem~\ref{thm:bgs}. It would be
interesting to have an algorithm that computes
the recovery sets of Theorem~\ref{thm:main}\ref{main:a} in time polynomial in the number $q^s-1$
of servers.

\emph{The list size $N+1$.} Open Problem~3.13 of~\cite{KRS} asks whether
$\FB(s,N+1,q)=2N$ for all $q$ and $s$. This equality is known for $s\le2$~\cite[Section~3]{KRS} and for
$q=2$~\cite{HKRS,KRS,YY}, where it is the case $m=2$ of
Corollary~\ref{cor:length}\ref{length:a}; Corollary~\ref{cor:length}\ref{length:b} establishes it
for odd $q$. We do not know whether it holds for even $q\ge4$ and $s\ge3$. In these cases
Lemma~\ref{lem:counting} gives $\FB(s,N+1,q)\ge2N$, and equality would follow if two nonzero
vectors on each line of $\F_q^s$ served every list of $N+1$ requests, as they do for odd $q$ by
Theorem~\ref{thm:tworep}.

\begin{appendices}

\section[Perfect matchings of \texorpdfstring{$\F_2^n$}{F\_2\textasciicircum n} with prescribed differences]{Perfect matchings of $\F_2^n$ with prescribed\\ differences}\label{app:bgs}

Theorem~\ref{thm:bgs} is proved in~\cite{Bin}. This appendix contains a complete proof, which
follows~\cite{Bin}. Throughout, $n\ge2$, $G\coloneqq\F_2^n$,
$M\coloneqq2^{n-1}$, so that $|G|=2M$, and $d_1,\dots,d_M$ are nonzero elements of $G$ with
$d_1+\dots+d_M=0$. A \emph{realization} of the list is a pair $(\mu,\varphi)$ that consists of a
perfect matching $\mu$ of $G$, that is, a partition of $G$ into pairs, and of a bijection
$\varphi$ from $\mu$ to $\{1,\dots,M\}$ such that $x+x'=d_{\varphi(\{x,x'\})}$ for every pair
$\{x,x'\}\in\mu$. Theorem~\ref{thm:bgs} states that a realization exists.

The proof assigns a sign to every realization and shows that the sum of these signs is an integer
which is divisible by $2^{M-1}$ and not by $2^M$; in particular it is not zero.

The realizations are encoded in a matrix. For a nonzero $d\in G$ we use the permutation matrix of
the translation $x\mapsto x+d$, with the sign $(-1)^{\eta_d(x)}$ in row $x$, where $\eta_d$ is a
linear functional with $\eta_d(d)=1$. We call this matrix $J_d$. The condition $\eta_d(d)=1$ makes
$J_d$ skew-symmetric. Hence the matrix $J_0+\sum_iy_iJ_{d_i}$, with indeterminates $y_i$ and one
fixed matrix $J_0$ of the same kind, has a Pfaffian; the term $J_0$ gives this Pfaffian the
constant term $\pm1$, and the coefficient of $y_1\cdots y_M$ in
it is the signed sum over the realizations (Subsection~\ref{subsec:pfaffian}). The
square of the Pfaffian is a determinant, and multiplication by $J_0^{-1}$ brings it to the form
$\det(\mathrm{Id}+\sum_iy_iK_i)$, where $K_i$ denotes the matrix $J_0^{-1}J_{d_i}$. The logarithm
of the Pfaffian, taken with the sign that makes the constant term of the Pfaffian equal to $1$,
therefore has coefficients that are rational linear combinations of traces of products of the
matrices $K_i$ (Subsection~\ref{subsec:log}). Such a product is, up to its sign, again a signed
translation, and a bilinear form $\omega$ decides whether two matrices of this kind commute or
anticommute (Subsection~\ref{subsec:weyl}). We group the orderings of the factors of
these products into equivalence classes, called swap classes, within which the products either
cancel or are all equal. This shows that, in the logarithm, the coefficient of a product of
$u<M$ distinct indeterminates is zero or has $2$-adic valuation at least $u$, and that the
coefficient of $y_1\cdots y_M$ is $2^{M-1}$ times a sum of signs
(Subsection~\ref{subsec:trees}). The number of these signs is the number of certain trees, and
we show that it is odd (Subsection~\ref{subsec:parity}). By these bounds, the coefficient of
$y_1\cdots y_M$ in the Pfaffian itself, which is obtained by exponentiating, then has $2$-adic
valuation $M-1$.

The functionals $\eta_d$ are defined by means of a basis of $G$ that is adapted to the list and
is constructed in Subsection~\ref{subsec:flag}. Together with the hypothesis
$d_1+\dots+d_M=0$, this choice makes the product of all the matrices $K_i$ equal to
$\pm\mathrm{Id}$, so that its trace is $\pm2M$. The parity argument of
Subsection~\ref{subsec:parity} also uses this basis.

The notation differs from that of~\cite{Bin}; the main differences are the following. The group
$G$ and the indeterminates $y_i$ are denoted there by $V$ and $t_i$. The subspaces $F_j$, the level $\hgt(d)$, the form $\omega$, the
matrices $K_i$ and their classes $\kappa_i$, the polynomial $\Phi$ and the coefficients
$\Lambda_S$, all of which are defined below, are denoted there by $H_j$, $h(d)$,
$\langle\cdot,\cdot\rangle$, $B_i$, $b_i$, $F$ and $L_S$, and the swap classes and pairing trees
correspond to the forests and trees of~\cite{Bin}.

\subsection{A basis adapted to the list}\label{subsec:flag}

\begin{lemma}\label{lem:flag}
There is a basis $e_1,\dots,e_n$ of $G$ that is adapted to the list in the following sense: for
$0\le j\le n$ let $F_j$
be the span of $e_1,\dots,e_j$; then for every $j\ge1$ the number of indices $i$ with
$d_i\in F_j\setminus F_{j-1}$ is even.
\end{lemma}

\begin{proof}
We construct $e_1,\dots,e_n$ one after the other. Suppose that $e_1,\dots,e_{j-1}$ have been
chosen, where $1\le j\le n$, and that the property holds for every $j'<j$. The
number of indices $i$ with $d_i\in F_{j-1}$ is then even, because $F_{j-1}\setminus\{0\}$ is the
union of the sets $F_k\setminus F_{k-1}$ with $k<j$ and no $d_i$ is zero. Since $M$ is even, the
number of indices $i$ with $d_i\notin F_{j-1}$ is even as well. These indices are distributed
over the cosets of $F_{j-1}$ different from $F_{j-1}$, and the number of such cosets is
$2^{n-j+1}-1$, which is odd. Hence some coset $x+F_{j-1}\ne F_{j-1}$ contains $d_i$ for an even
number of indices $i$. We put $e_j\coloneqq x$; then $F_j\setminus F_{j-1}=x+F_{j-1}$.
\end{proof}

We fix a basis as in Lemma~\ref{lem:flag} and let $e_1^{*},\dots,e_n^{*}$ be the dual basis of
the space $G^{*}$ of linear functionals $G\to\F_2$. The \emph{level} $\hgt(d)$ of a nonzero
$d\in G$ is the index $j$ with $d\in F_j\setminus F_{j-1}$, and we put
\[
  \eta_d\coloneqq e^{*}_{\hgt(d)}\in G^{*}.
\]
The functional $e_j^{*}$ vanishes on $F_{j-1}$ and takes the value $1$ on $F_j\setminus F_{j-1}$.
We shall use the following three properties.
\begin{enumerate}[label=\textup{(B\arabic*)}]
\item\label{H1} $\eta_d(d)=1$ for every nonzero $d\in G$.
\item\label{H2} $\eta_{d_1}+\dots+\eta_{d_M}=0$, because every $e_j^{*}$ occurs an even number of
times in this sum.
\item\label{H3} If $d,d'$ are nonzero elements of the same level $j$, then
$d+d'\in F_{j-1}$, and $\eta_{d''}$ vanishes on $F_{j-1}$ for every nonzero $d''$ of level at
least $j$.
\end{enumerate}

\subsection{Signed translations}\label{subsec:weyl}

The matrices of the overview are, up to sign, signed translations in the following sense; this
subsection
records how such operators multiply and commute and what their traces are.
Let $\R^G$ be the space of real functions on $G$. We identify a linear operator on $\R^G$ with
its matrix in the basis of indicator functions of the elements of $G$, so that
$(Tf)(x)=\sum_{y\in G}T_{x,y}f(y)$. For $a\in G$ and $\lambda\in G^{*}$ let $W(a,\lambda)$ be the
\emph{signed translation}
\[
  \bigl(W(a,\lambda)f\bigr)(x)\coloneqq(-1)^{\lambda(x)}f(x+a),
  \qquad\text{that is,}\qquad
  W(a,\lambda)_{x,y}=\begin{cases}(-1)^{\lambda(x)}&\text{if }y=x+a,\\0&\text{otherwise.}\end{cases}
\]
On the group $G\times G^{*}$ we consider the form
\[
  \omega\bigl((a,\lambda),(b,\nu)\bigr)\coloneqq\lambda(b)+\nu(a)\in\F_2 .
\]
It is bilinear and symmetric, and $\omega(\gamma,\gamma)=0$ for every $\gamma\in G\times G^{*}$.

\begin{lemma}\label{lem:weyl}
Let $a,b\in G$ and $\lambda,\nu\in G^{*}$.
\begin{enumerate}[label=\textup{(\alph*)}]
\item\label{weyl:prod} $W(a,\lambda)W(b,\nu)=(-1)^{\nu(a)}W(a+b,\lambda+\nu)$.
\item\label{weyl:comm} $W(a,\lambda)W(b,\nu)=(-1)^{\omega((a,\lambda),(b,\nu))}\,W(b,\nu)W(a,\lambda)$.
\item\label{weyl:transpose} The transpose of $W(a,\lambda)$ is $(-1)^{\lambda(a)}W(a,\lambda)$, and
$W(a,\lambda)^2=(-1)^{\lambda(a)}\,\mathrm{Id}$.
\item\label{weyl:trace} $\tr W(a,\lambda)=2M$ if $(a,\lambda)=(0,0)$, and $\tr W(a,\lambda)=0$
otherwise.
\end{enumerate}
\end{lemma}

\begin{proof}
\ref{weyl:prod} For $f\in\R^G$ and $x\in G$,
\[
  \bigl(W(a,\lambda)W(b,\nu)f\bigr)(x)=(-1)^{\lambda(x)}(-1)^{\nu(x+a)}f(x+a+b)
  =(-1)^{\nu(a)}(-1)^{(\lambda+\nu)(x)}f(x+a+b).
\]
\ref{weyl:comm} follows from~\ref{weyl:prod} applied to both products. \ref{weyl:transpose} The
$(x,y)$ entry of the transpose is $W(a,\lambda)_{y,x}$, which is $(-1)^{\lambda(y)}$ if $x=y+a$
and $0$ otherwise; for $y=x+a$ we have $(-1)^{\lambda(y)}=(-1)^{\lambda(a)}(-1)^{\lambda(x)}$. The
second assertion is the case $(b,\nu)=(a,\lambda)$ of~\ref{weyl:prod}. \ref{weyl:trace} The
diagonal entries vanish unless $a=0$, and $\tr W(0,\lambda)=\sum_{x\in G}(-1)^{\lambda(x)}$, which
is $2M$ for $\lambda=0$ and $0$ otherwise.
\end{proof}

We say that an operator of the form $\pm W(a,\lambda)$ has \emph{class} $(a,\lambda)\in G\times
G^{*}$. By Lemma~\ref{lem:weyl}\ref{weyl:prod} a product of operators with classes
$\gamma_1,\dots,\gamma_k$ has class $\gamma_1+\dots+\gamma_k$, whatever the order of the factors.
By~\ref{weyl:comm} two such operators commute if the value of $\omega$ on their classes is $0$,
and anticommute if it is $1$. By~\ref{weyl:trace} an operator of class $(0,0)$ has trace $\pm2M$,
and an operator of any other class has trace $0$.

For a nonzero $d\in G$ let
\[
  J_d\coloneqq W(d,\eta_d),\qquad\text{and let}\qquad J_0\coloneqq J_{e_n}=W(e_n,e_n^{*}).
\]
By~\ref{H1} and Lemma~\ref{lem:weyl}\ref{weyl:transpose} the matrices $J_d$ are skew-symmetric
and satisfy $J_d^2=-\mathrm{Id}$. In particular $J_0^{-1}=-J_0$, and we put
\[
  K_i\coloneqq J_0^{-1}J_{d_i}\quad(1\le i\le M),\qquad\text{an operator of class}\quad
  \kappa_i\coloneqq\bigl(e_n+d_i,\ e_n^{*}+\eta_{d_i}\bigr).
\]
The subscript $0$ of $J_0$ is only a label. The matrix $J_0$ gives the Pfaffian of
Subsection~\ref{subsec:pfaffian} the constant term $\pm1$, without which the logarithm of
Subsection~\ref{subsec:log} could not be taken; the choice of $e_n$, whose level $n$ is the
largest possible, is used in Step~1 of the proof of Lemma~\ref{lem:parity}.

\subsection{The Pfaffian and the realizations}\label{subsec:pfaffian}

Fix a total order on $G$. For a skew-symmetric matrix $Y=(Y_{x,x'})_{x,x'\in G}$ with entries in
$\Z$ or in a polynomial ring over $\Z$, the Pfaffian is
\[
  \Pf(Y)\coloneqq\sum_{\mu}\sgn(\mu)\prod_{\{x,x'\}\in\mu,\ x<x'}Y_{x,x'},
\]
where the sum is over the perfect matchings $\mu$ of $G$. Here $\sgn(\mu)$ is the sign of the
permutation that maps the elements of $G$, in increasing order, to the sequence
$x_1,x_1',x_2,x_2',\dots,x_M,x_M'$, where $\{x_k,x_k'\}$, $1\le k\le M$, are the pairs of $\mu$,
numbered in any way, and $x_k<x_k'$; this sign does not depend on the numbering of the pairs. We
use two standard facts, for which we refer
to~\cite{LP}: the Pfaffian is a polynomial with integer coefficients in the entries of $Y$, and
$\Pf(Y)^2=\det Y$.

Let $y_1,\dots,y_M$ be independent indeterminates and put
\[
  \Psi\coloneqq\Pf\Bigl(J_0+\sum_{i=1}^My_iJ_{d_i}\Bigr)\in\Z[y_1,\dots,y_M].
\]

\begin{lemma}\label{lem:realizations}
The coefficient of the monomial $y_1y_2\cdots y_M$ in $\Psi$ is a sum of signs
$\varepsilon(\mu,\varphi)=\pm1$, one for each realization $(\mu,\varphi)$ of the list
$d_1,\dots,d_M$. In particular, if this coefficient is not zero, then a realization exists.
\end{lemma}

\begin{proof}
For $x<x'$ the $(x,x')$ entry of $J_0+\sum_iy_iJ_{d_i}$ is the sum of a constant, which is $\pm1$
or $0$, and of the terms $(-1)^{\eta_{d_i}(x)}y_i$ for the indices $i$ with $d_i=x+x'$. A perfect
matching has $M$ pairs. In the expansion of the product over the pairs of $\mu$, the monomial
$y_1\cdots y_M$ arises exactly from the choices of one term $\pm y_i$ in each pair such that the
$M$ chosen indices are distinct; these choices are the bijections $\varphi$ for which
$(\mu,\varphi)$ is a realization, and each of them contributes $\pm\sgn(\mu)$.
\end{proof}

Since $J_0^2=-\mathrm{Id}$, we have $(\det J_0)^2=\det(-\mathrm{Id})=1$, and
$\det J_0=\Pf(J_0)^2\ge0$; hence $\det J_0=1$ and $\varepsilon_0\coloneqq\Pf(J_0)=\pm1$. Put
\[
  \Phi\coloneqq\varepsilon_0\Psi .
\]
Then $\Phi\in\Z[y_1,\dots,y_M]$ has constant term $\varepsilon_0\Pf(J_0)=1$, and
\begin{equation}\label{eq:square}
  \Phi^2=\det\Bigl(J_0+\sum_iy_iJ_{d_i}\Bigr)=\det(J_0)\,\det\Bigl(\mathrm{Id}+\sum_iy_iK_i\Bigr)
  =\det\Bigl(\mathrm{Id}+\sum_{i=1}^My_iK_i\Bigr).
\end{equation}
By Lemma~\ref{lem:realizations}, Theorem~\ref{thm:bgs} follows once we show that the coefficient
of $y_1\cdots y_M$ in $\Phi$ is not zero. We shall prove that its $2$-adic valuation is exactly
$M-1$.

\subsection{The logarithm}\label{subsec:log}

Only the coefficient of $y_1\cdots y_M$ matters, so we work in the ring
\[
  \cA\coloneqq\Q[y_1,\dots,y_M]/(y_1^2,\dots,y_M^2).
\]
The monomials $y_S\coloneqq\prod_{i\in S}y_i$, $S\subseteq\{1,\dots,M\}$, form a basis of $\cA$
over $\Q$, and $y_Sy_{S'}$ equals $y_{S\cup S'}$ if $S\cap S'=\emptyset$ and is $0$ otherwise. The
coefficient of $y_1\cdots y_M$ in a polynomial is unchanged when the polynomial is replaced by
its image in $\cA$. Let $\fm$ be the ideal of $\cA$ spanned by the $y_S$ with $S\ne\emptyset$; it
satisfies $\fm^{M+1}=0$. The series
$\exp(g)=\sum_{k\ge0}g^k/k!$ and $\log(1+g)=\sum_{k\ge1}(-1)^{k-1}g^k/k$ are therefore finite sums
for $g\in\fm$. They define mutually inverse bijections between $\fm$ and $1+\fm$, and
$\log(fg)=\log f+\log g$ for $f,g\in1+\fm$. For a square matrix $Y$ with entries in $\fm$ we
have
\begin{equation}\label{eq:logdet}
  \log\det(\mathrm{Id}+Y)=\sum_{k\ge1}\frac{(-1)^{k-1}}{k}\,\tr(Y^k),
\end{equation}
which is the specialization of the corresponding identity of formal power series in the entries
of a matrix (equivalently, of $\det\exp Z=\exp\tr Z$ for $Z=\log(\mathrm{Id}+Y)$).

The image of $\Phi$ in $\cA$ lies in $1+\fm$. Let
\[
  \Lambda\coloneqq\log\Phi=\sum_{\emptyset\ne S\subseteq\{1,\dots,M\}}\Lambda_S\,y_S ,
  \qquad\Lambda_S\in\Q .
\]
For a nonempty set $S$ with $u$ elements, an \emph{ordering} of $S$ is a bijection
$\pi\colon\{1,\dots,u\}\to S$, and we write $K_{\pi}\coloneqq K_{\pi(1)}K_{\pi(2)}\cdots
K_{\pi(u)}$.

\begin{lemma}\label{lem:log}
\begin{enumerate}[label=\textup{(\alph*)}]
\item\label{log:a} For every nonempty $S\subseteq\{1,\dots,M\}$ with $u$ elements,
\[
  \Lambda_S=\frac{(-1)^{u-1}}{2u}\sum_{\pi}\tr(K_{\pi}),
\]
where the sum is over the $u!$ orderings $\pi$ of $S$.
\item\label{log:b} The coefficient of $y_1\cdots y_M$ in $\Phi$ equals
$\sum\prod_{j=1}^k\Lambda_{S_j}$, where the sum is over all partitions $\{S_1,\dots,S_k\}$ of
$\{1,\dots,M\}$ into nonempty sets.
\end{enumerate}
\end{lemma}

\begin{proof}
\ref{log:a} By~\eqref{eq:square} and~\eqref{eq:logdet} we have
$2\Lambda=\log(\Phi^2)=\sum_{k\ge1}\frac{(-1)^{k-1}}{k}\tr(Y^k)$ with $Y\coloneqq\sum_iy_iK_i$. Now
$Y^k=\sum y_{i_1}\cdots y_{i_k}K_{i_1}\cdots K_{i_k}$, summed over all sequences
$(i_1,\dots,i_k)$, and in $\cA$ the product $y_{i_1}\cdots y_{i_k}$ equals $y_S$ exactly when
$k=u$ and $(i_1,\dots,i_u)$ lists the elements of $S$ without repetition. Hence the coefficient
of $y_S$ in $2\Lambda$ is $\frac{(-1)^{u-1}}{u}\sum_{\pi}\tr(K_{\pi})$.

\ref{log:b} We have $\Phi=\exp(\Lambda)=\sum_{k\ge0}\Lambda^k/k!$. The coefficient of
$y_1\cdots y_M$ in $\Lambda^k$ is the sum of $\prod_j\Lambda_{S_j}$ over all sequences
$(S_1,\dots,S_k)$ of pairwise disjoint nonempty sets with union $\{1,\dots,M\}$, and every
partition into $k$ sets gives $k!$ such sequences.
\end{proof}

\subsection{Orderings, swap classes and valuations}\label{subsec:trees}

To bound the $2$-adic valuations of the numbers $\Lambda_S$ we group the orderings of $S$
into equivalence classes, the swap classes, on which the sum of the products $K_{\pi}$ can be
evaluated. For a nonzero rational
number $a$ let $v_2(a)$ be its $2$-adic valuation. For an integer
$u\ge0$ let $s_2(u)$ be the number of digits $1$ in the binary expansion of $u$; by Legendre's
formula, $v_2(u!)=u-s_2(u)$.

We call a finite sequence of distinct elements of $\{1,\dots,M\}$ a \emph{word}; the
concatenation of words $w,w'$ with no common element is written $ww'$. For a word
$w=(i_1,\dots,i_k)$ let $K_w\coloneqq K_{i_1}\cdots K_{i_k}$, an operator of class
$\kappa(w)\coloneqq\kappa_{i_1}+\dots+\kappa_{i_k}$. For a word $w$ whose length is a power of
two we define a set $\Sw(w)$ of words and the property of being \emph{good} by recursion on the
length. If $w$ has length $1$, then $\Sw(w)\coloneqq\{w\}$ and $w$ is good. If $w$ has length
$2^k\ge2$, let $w=w'w''$ with $w'$ and $w''$ of length $2^{k-1}$; then
\[
  \Sw(w)\coloneqq\{\,v'v'',\ v''v':\ v'\in\Sw(w'),\ v''\in\Sw(w'')\,\},
\]
and $w$ is good if $w'$ and $w''$ are good and $\omega(\kappa(w'),\kappa(w''))=0$, that is, if
$K_{w'}$ and $K_{w''}$ commute (Lemma~\ref{lem:weyl}\ref{weyl:comm}). Thus
$\Sw(w)$ consists of the words obtained from $w$ by exchanging or keeping, independently, the two
halves of $w$, the two halves of each half, and so on.

\begin{lemma}\label{lem:swap}
Let $w$ be a word of length $2^k$.
\begin{enumerate}[label=\textup{(\alph*)}]
\item\label{swap:size} $|\Sw(w)|=2^{2^k-1}$, and $\Sw(v)=\Sw(w)$ for every $v\in\Sw(w)$.
\item\label{swap:sum} $\sum_{v\in\Sw(w)}K_v=2^{2^k-1}K_w$ if $w$ is good, and
$\sum_{v\in\Sw(w)}K_v=0$ otherwise.
\end{enumerate}
\end{lemma}

\begin{proof}
Both parts are proved by induction on $k$, the case $k=0$ being clear. Let $k\ge1$ and
$w=w'w''$ as above.

\ref{swap:size} The words $v'v''$ and $v''v'$ with $v'\in\Sw(w')$, $v''\in\Sw(w'')$ are pairwise
distinct, because $w'$ and $w''$ have no common element. Hence
$|\Sw(w)|=2\,|\Sw(w')|\,|\Sw(w'')|=2\cdot2^{2^{k-1}-1}\cdot2^{2^{k-1}-1}=2^{2^k-1}$. If
$v'\in\Sw(w')$ and $v''\in\Sw(w'')$, then $\Sw(v')=\Sw(w')$ and $\Sw(v'')=\Sw(w'')$ by induction,
and the definition gives $\Sw(v'v'')=\Sw(v''v')=\Sw(w)$.

\ref{swap:sum} Write $\Sigma(w)\coloneqq\sum_{v\in\Sw(w)}K_v$. By the definition of $\Sw(w)$,
\[
  \Sigma(w)=\Sigma(w')\Sigma(w'')+\Sigma(w'')\Sigma(w').
\]
If $w'$ or $w''$ is not good, then $\Sigma(w)=0$ by induction, and $w$ is not good. If both are
good, then by induction and Lemma~\ref{lem:weyl}\ref{weyl:comm}
\[
  \Sigma(w)=2^{2^k-2}\bigl(K_{w'}K_{w''}+K_{w''}K_{w'}\bigr)
  =2^{2^k-2}\bigl(1+(-1)^{\omega(\kappa(w'),\kappa(w''))}\bigr)K_{w'}K_{w''},
\]
and $K_{w'}K_{w''}=K_w$. This is $2^{2^k-1}K_w$ if $\omega(\kappa(w'),\kappa(w''))=0$, and $0$
otherwise.
\end{proof}

Now let $S$ be a set with $u\ge1$ elements, and write $u=2^{k_1}+\dots+2^{k_r}$ with
$k_1>\dots>k_r\ge0$, so that $r=s_2(u)$. We cut every ordering $\pi$ of $S$, regarded as the word
$(\pi(1),\dots,\pi(u))$, into consecutive \emph{blocks} $\pi^{(1)},\dots,\pi^{(r)}$ of lengths
$2^{k_1},\dots,2^{k_r}$. Two orderings $\pi,\pi'$ are \emph{equivalent} if
$\pi'^{(j)}\in\Sw(\pi^{(j)})$ for all $j$; by Lemma~\ref{lem:swap}\ref{swap:size} this is an
equivalence relation. We call its equivalence classes \emph{swap classes}; every swap class has
$\prod_j2^{2^{k_j}-1}=2^{u-s_2(u)}=2^{v_2(u!)}$ elements, the largest power of $2$ that divides
the number $u!$ of orderings. We call $\pi$ good if all its blocks are good. Since
$K_{\pi}=K_{\pi^{(1)}}\cdots K_{\pi^{(r)}}$, Lemma~\ref{lem:swap}\ref{swap:sum} shows that the sum
of $K_{\pi'}$ over the orderings $\pi'$ equivalent to $\pi$ is $2^{u-s_2(u)}K_{\pi}$ if $\pi$ is
good, and $0$ otherwise. In particular, an ordering that is equivalent to a good ordering is
good, because the two orderings have the same swap class and the sum over this class is not
zero. Summing over the swap classes and using
Lemma~\ref{lem:log}\ref{log:a},
we obtain
\begin{equation}\label{eq:LambdaS}
  \Lambda_S=\frac{(-1)^{u-1}}{2u}\;2^{u-s_2(u)}\sum_{\pi}\tr(K_{\pi}),
\end{equation}
where now the sum is over a set of good orderings of $S$ that contains exactly one ordering from
each swap class of good orderings.

\begin{lemma}\label{lem:valuations}
\begin{enumerate}[label=\textup{(\alph*)}]
\item\label{val:small} If $S\subseteq\{1,\dots,M\}$ and $1\le|S|<M$, then $\Lambda_S=0$ or
$v_2(\Lambda_S)\ge|S|$.
\item\label{val:top} $\Lambda_{\{1,\dots,M\}}=2^{M-1}\,E$, where $E$ is a sum of signs $\pm1$, one
for each swap class of good orderings of $\{1,\dots,M\}$.
\end{enumerate}
\end{lemma}

\begin{proof}
Each $K_{\pi}$ in~\eqref{eq:LambdaS} is an operator of class
$\kappa(S)\coloneqq\sum_{i\in S}\kappa_i$, so, by Lemma~\ref{lem:weyl}\ref{weyl:trace}, its trace is $0$ or $\pm2M=\pm2^n$.

\ref{val:small} Let $u\coloneqq|S|$. By~\eqref{eq:LambdaS}, if $\Lambda_S\ne0$ then
\[
  v_2(\Lambda_S)\ \ge\ -1-v_2(u)+\bigl(u-s_2(u)\bigr)+n
  =n-1+v_2\bigl((u-1)!\bigr)=n-1+(u-1)-s_2(u-1),
\]
where we used $v_2(u!)=u-s_2(u)$ twice. Since $0\le u-1\le2^{n-1}-2$, the number $u-1$ is smaller than
$2^{n-1}-1$, which is the only nonnegative integer below $2^{n-1}$ with $n-1$ binary digits
equal to $1$; thus $s_2(u-1)\le n-2$, and $v_2(\Lambda_S)\ge u$.

\ref{val:top} Here $u=M=2^{n-1}$, $s_2(M)=1$, and every ordering has a single block. The class of
every $K_{\pi}$ is
\[
  \sum_{i=1}^M\kappa_i=\Bigl(Me_n+\sum_id_i,\ Me_n^{*}+\sum_i\eta_{d_i}\Bigr)=(0,0),
\]
because $M$ is even, $\sum_id_i=0$, and $\sum_i\eta_{d_i}=0$ by~\ref{H2}. Hence
$\tr(K_{\pi})=\pm2M$ for every ordering, and~\eqref{eq:LambdaS} gives
$\Lambda_{\{1,\dots,M\}}=\frac{(-1)^{M-1}}{2M}\,2^{M-1}\sum_{\pi}(\pm2M)=2^{M-1}E$.
\end{proof}

\subsection{The parity of the number of good pairing trees}\label{subsec:parity}

It remains to show that the number of swap classes of good orderings of $\{1,\dots,M\}$ is
odd. We describe the swap classes by trees. A \emph{pairing tree} on a set $\Omega$ whose size
$2^k$ is a power of two is a sequence $(\fB_0,\fB_1,\dots,\fB_k)$ of partitions of $\Omega$ such
that $\fB_0$ consists of the one-element subsets, $\fB_k=\{\Omega\}$, and every member of $\fB_j$,
$j\ge1$, is the union of exactly two members of $\fB_{j-1}$. The members of the partitions are
the \emph{nodes} of the tree, and the two members of $\fB_{j-1}$ contained in a node
$P\in\fB_j$, $j\ge1$, are the \emph{children} of $P$.

Words, the sets $\Sw$ and swap classes are defined in the same way for any finite set in place of
$\{1,\dots,M\}$. An ordering $\pi$ of $\Omega$ determines a pairing tree: $\fB_j$ consists of the
sets of entries of the $2^{k-j}$ subwords of length $2^j$ into which the word $\pi$ is cut. Two orderings determine the
same pairing tree if and only if they are equivalent. Indeed, the two halves of a word and the two
halves of any word in its set $\Sw$ have the same sets of entries, in some order, which gives
one direction by induction; conversely, if two words have the same pairing tree, then their
halves have the same sets of entries, in some order, and the corresponding halves have the same
pairing trees, so the claim again follows by induction. Every pairing tree
comes from an ordering, obtained by listing the elements of the two children of each node one after
the other. Since every swap class has $2^{2^k-1}$ elements, the number of pairing trees on
$\Omega$ is
\begin{equation}\label{eq:numtrees}
  \frac{(2^k)!}{2^{2^k-1}},\qquad\text{which is odd, because }v_2\bigl((2^k)!\bigr)=2^k-1 .
\end{equation}

Let $\Omega=\{1,\dots,M\}$. For a node $P$ put $\kappa(P)\coloneqq\sum_{i\in P}\kappa_i$. A
pairing tree is \emph{good} if $\omega(\kappa(P'),\kappa(P''))=0$ for the two children $P',P''$
of every node with at least two elements. By the definition of good words, an ordering is good if
and only if its pairing tree is good. The number $E$ in Lemma~\ref{lem:valuations}\ref{val:top}
is therefore a sum of signs, one for each good pairing tree on $\{1,\dots,M\}$.

\begin{lemma}\label{lem:parity}
The number of good pairing trees on $\{1,\dots,M\}$ is odd.
\end{lemma}

\begin{proof}
By Lemma~\ref{lem:flag}, for every $j$ the number of indices $i$ with $\hgt(d_i)=j$, that is, with
$d_i\in F_j\setminus F_{j-1}$, is even. We
can therefore partition $\{1,\dots,M\}$ into $M/2$ pairs $p_1,\dots,p_{M/2}$ such that the two
indices $i,i'$ of every pair satisfy $\hgt(d_i)=\hgt(d_{i'})$; we call this common value the
level of the pair, and we number the pairs so that their levels are nondecreasing. For
$0\le k\le M/2$ let $\cT_k$ be the set of good pairing trees in which each of the pairs
$p_1,\dots,p_k$ is a node, that is, a member of $\fB_1$. Thus $\cT_0$ is the set of all good
pairing trees. We show that
\begin{equation}\label{eq:paritystep}
  |\cT_{k-1}|\equiv|\cT_k|\pmod2\qquad(1\le k\le M/2)
\end{equation}
and that $|\cT_{M/2}|$ is odd.

Fix $k$, let $p_k=\{i,i'\}$, let $j$ be the level of $p_k$, and put $\delta\coloneqq
d_i+d_{i'}$. By~\ref{H3}, $\delta\in F_{j-1}$, and since $\eta_{d_i}=\eta_{d_{i'}}=e_j^{*}$ and
$\kappa_i=(e_n+d_i,\,e_n^{*}+\eta_{d_i})$, we have
$\kappa_i+\kappa_{i'}=(\delta,0)$. We write $\Delta\coloneqq(\delta,0)$ and note that
$\omega(\Delta,(b,\nu))=\nu(\delta)$ for all $(b,\nu)\in G\times G^{*}$. We
prove~\eqref{eq:paritystep} by means of an involution: the exchange of $i$ and $i'$ maps
$\cT_{k-1}$ to itself (Steps~1--3), and the trees that it fixes are exactly those of $\cT_k$
(Step~4). Step~5 shows that $|\cT_{M/2}|$ is odd.

\emph{Step 1.} If $l$ is an index that belongs to none of the pairs $p_1,\dots,p_{k-1}$, then
$\omega(\Delta,\kappa_l)=0$. Indeed, $l$ belongs to a pair $p_{k'}$ with $k'\ge k$, so
$\hgt(d_l)\ge j$, and $\omega(\Delta,\kappa_l)=e_n^{*}(\delta)+\eta_{d_l}(\delta)=0$ by~\ref{H3},
applied to $d''=e_n$ and to $d''=d_l$. If $\{l,l'\}$ is one of the pairs $p_1,\dots,p_{k-1}$,
then $\kappa_l+\kappa_{l'}=(d_l+d_{l'},0)$ and $\omega(\Delta,\kappa_l+\kappa_{l'})=0$.

\emph{Step 2.} Let $\tau$ be a tree in $\cT_{k-1}$ and let $P$ be a node of $\tau$ that has at
least two elements, or that consists of one index outside the pairs $p_1,\dots,p_{k-1}$. Then
$\omega(\Delta,\kappa(P))=0$. For a node with one element this is Step~1. A node with at least
two elements is a union of members of $\fB_1$, and the pairs $p_1,\dots,p_{k-1}$ are members of
$\fB_1$; hence the node is a disjoint union of some of these pairs and of indices outside them,
and the assertion follows from Step~1 because $\omega$ is bilinear.

\emph{Step 3.} Let $\theta$ be the permutation of $\{1,\dots,M\}$ that exchanges $i$ and $i'$
and fixes all other indices. For a pairing tree $\tau=(\fB_0,\dots,\fB_{n-1})$ let $\theta\tau$
be the pairing tree whose partitions consist of the sets $\theta(P)$, $P\in\fB_j$. We claim that
$\theta$ maps $\cT_{k-1}$ to itself. The pairs $p_1,\dots,p_{k-1}$ contain neither $i$ nor $i'$,
so they are nodes of $\theta\tau$ if they are nodes of $\tau$. It remains to show that
$\theta\tau$ is good if $\tau\in\cT_{k-1}$. For a node $P$ of $\tau$ we have
$\kappa(\theta(P))=\kappa(P)$ if $P$ contains both or none of $i,i'$, and
$\kappa(\theta(P))=\kappa(P)+\Delta$ if $P$ contains exactly one of them. Let $P'$ and $P''$ be
the children of a node of $\tau$; we must show that
$\omega(\kappa(\theta(P')),\kappa(\theta(P'')))=\omega(\kappa(P'),\kappa(P''))$.
\begin{itemize}
\item If each of $P',P''$ contains both or none of $i,i'$, there is nothing to prove.
\item Suppose that $P'$ contains exactly one of $i,i'$ and $P''$ contains none. The difference of
the two values of $\omega$ is $\omega(\Delta,\kappa(P''))$. If $P''$ has at least two elements,
this is $0$ by Step~2. If $P''=\{l\}$, then $P'$ is $\{i\}$ or $\{i'\}$, and $P'\cup P''$ is the
member of $\fB_1$ that contains $l$. It contains an index of $p_k$, so it is none of the pairs
$p_1,\dots,p_{k-1}$; since these pairs are members of $\fB_1$, the index $l$ lies outside them,
and Step~2 applies again.
\item Suppose that $P'$ contains one of $i,i'$ and $P''$ contains the other. The difference of
the two values is
$\omega(\kappa(P'),\Delta)+\omega(\Delta,\kappa(P''))+\omega(\Delta,\Delta)$. The last term is
$0$, and the first two are $0$ by Step~2, which applies to $P'$ and $P''$ because the indices $i$
and $i'$ lie outside the pairs $p_1,\dots,p_{k-1}$.
\end{itemize}
The remaining cases are obtained by exchanging $P'$ and $P''$. This proves the claim.

\emph{Step 4.} The map $\tau\mapsto\theta\tau$ is an involution of $\cT_{k-1}$, and its fixed
points are exactly the trees in $\cT_k$. Indeed, if $\{i,i'\}$ is a node of $\tau$, then every
node with at least two elements contains both or none of $i,i'$, and $\theta\tau=\tau$.
Conversely, suppose that $\theta\tau=\tau$ and that the members of $\fB_1$ containing $i$ and
$i'$ are different, say $\{i,a\}$ and $\{i',b\}$. Then $\theta(\{i,a\})=\{i',a\}$ is a member of
$\fB_1$ that contains $i'$, so $a=b$, which is impossible because the two members are disjoint.
An involution of a finite set has a number of fixed points that is congruent modulo $2$ to the
size of the set, and~\eqref{eq:paritystep} follows.

\emph{Step 5.} A pairing tree in which all the pairs $p_1,\dots,p_{M/2}$ are nodes is good. For a
node $p_k=\{i,i'\}$ we have, with $\Delta=\kappa_i+\kappa_{i'}$ as above,
$\omega(\kappa_i,\kappa_{i'})=\omega(\kappa_i,\kappa_i)+\omega(\kappa_i,\Delta)=0$ by Step~1. The
children $P',P''$ of a node with at least four elements are unions of pairs $p_k$, so
$\kappa(P')$ and $\kappa(P'')$ have the form $(b,0)$ and $(b',0)$, and
$\omega((b,0),(b',0))=0$. Hence $\cT_{M/2}$ is the set of all
pairing trees with $\fB_1=\{p_1,\dots,p_{M/2}\}$. These correspond to the pairing trees on the
set of the $M/2=2^{n-2}$ pairs, and their number is odd by~\eqref{eq:numtrees}.
\end{proof}

\begin{proof}[Proof of Theorem~\ref{thm:bgs}]
By Lemma~\ref{lem:log}\ref{log:b} the coefficient of $y_1\cdots y_M$ in $\Phi$ is the sum of the
products $\prod_j\Lambda_{S_j}$ over the partitions of $\{1,\dots,M\}$. The partition with the
single member $\{1,\dots,M\}$ contributes $2^{M-1}E$. By
Lemma~\ref{lem:valuations}\ref{val:top} and the paragraph before Lemma~\ref{lem:parity}, $E$ is
a sum of signs, one for each good pairing tree on $\{1,\dots,M\}$, and the number of these trees
is odd by Lemma~\ref{lem:parity}; hence $E$ is odd and this contribution has $2$-adic valuation
$M-1$. In every other partition all members have fewer than
$M$ elements, and by Lemma~\ref{lem:valuations}\ref{val:small} the product $\prod_j\Lambda_{S_j}$
is $0$ or has valuation at least $\sum_j|S_j|=M$. Therefore the coefficient of $y_1\cdots y_M$ in
$\Phi$ has valuation exactly $M-1$. It is not zero, and by Lemma~\ref{lem:realizations} the list
$d_1,\dots,d_M$ has a realization.
\end{proof}

\end{appendices}

\pdfbookmark[1]{References}{references}

\end{document}